\documentclass[aps,prl,twocolumn,amsfonts,nobibnotes,superscriptaddress,showpacs]{revtex4}

\usepackage{dcolumn}
\usepackage{amsmath}
\usepackage{amssymb}
\usepackage{graphicx}
\usepackage{bm}
\usepackage{color}
\usepackage{titlesec}

\begin{document}

\title{Real-time Observation of Phonon-Mediated $\sigma$-$\pi$ Interband Scattering in MgB$_2$}

\vspace{2cm}

\author{E. Baldini}
	\affiliation{Institute of Physics, \'Ecole Polytechnique F\'{e}d\'{e}rale de Lausanne, CH-1015 Lausanne, Switzerland}
	\affiliation{Institute of Chemical Sciences and Engineering, \'Ecole Polytechnique F\'{e}d\'{e}rale de Lausanne, CH-1015 Lausanne, Switzerland}

\author{A. Mann}
	\affiliation{Institute of Physics, \'Ecole Polytechnique F\'{e}d\'{e}rale de Lausanne, CH-1015 Lausanne, Switzerland}
	
\author{L. Benfatto}
\affiliation{Institute for Complex Systems - CNR, and Physics Department, University of Rome ``La Sapienza", I-00185 Rome, Italy}
	
\author{E. Cappelluti}
\affiliation{Institute for Complex Systems - CNR, and Physics Department, University of Rome ``La Sapienza", I-00185 Rome, Italy}
	
\author{A. Acocella}
	\affiliation{Department of Chemistry ``G. Ciamician", Universit\`a di Bologna, I-40126, Bologna, Italy}
	
\author{V. M. Silkin}
	\affiliation{Departamento de F\'{\i}sica de Materiales, Universidad del Pa\'{\i}s Vasco, 20080 San Sebasti\'an/Donostia, Spain}
	\affiliation{Donostia International Physics Center, 20018 San Sebasti\'an/Donostia, Spain}
	\affiliation{IKERBASQUE, Basque Foundation for Science, 48011, Bilbao, Spain}
	
\author{S. V. Eremeev}
	\affiliation{Institute of Strength Physics and Materials Science, 634055, Tomsk, Russia}
	\affiliation{Tomsk State University, 634050, Tomsk, Russia}

\author{A. B. Kuzmenko}
	\affiliation{Department of Quantum Matter Physics, University of Geneva, CH-1211 Geneva 4, Switzerland}
	
\author{S. Borroni}
	\affiliation{Institute of Physics, \'Ecole Polytechnique F\'{e}d\'{e}rale de Lausanne, CH-1015 Lausanne, Switzerland}
	
\author{T. Tan}
	\affiliation{Department of Material Science and Engineering, The Pennsylvania State University, PA 16802, USA}
	
\author{X. X. Xi}
	\affiliation{Department of Material Science and Engineering, The Pennsylvania State University, PA 16802, USA}
	
\author{F. Zerbetto}
	\affiliation{Department of Chemistry ``G. Ciamician", Universit\`a di Bologna, I-40126, Bologna, Italy}
	
\author{R. Merlin}
	\affiliation{Department of Physics, Center for Photonics and Multiscale Nanomaterials, University of Michigan, Ann Arbor, Michigan 48109-1040, USA}
	
\author{F. Carbone}
	\affiliation{Institute of Physics, \'Ecole Polytechnique F\'{e}d\'{e}rale de Lausanne, CH-1015 Lausanne, Switzerland}

\date{\today}

\begin{abstract}
In systems having an anisotropic electronic structure, such as the layered materials graphite, graphene and cuprates, impulsive light excitation can coherently stimulate specific bosonic modes, with exotic consequences for the emergent electronic properties. Here we show that the population of $\mathrm{E_{2g}}$ phonons in the multiband superconductor MgB$_2$ can be selectively enhanced by femtosecond laser pulses, leading to a transient control of the number of carriers in the $\sigma$-electronic subsystem. The nonequilibrium evolution of the material optical constants is followed in the spectral region sensitive to both the a- and c-axis plasma frequencies and modeled theoretically, revealing the details of the $\sigma$-$\pi$ interband scattering mechanism in MgB$_2$.
\end{abstract}


\maketitle

Exotic phenomena and novel functionalities emerge in solids due to the coupling between the charge carriers and collective modes of structural, electronic and magnetic origin. Key to exploiting this phenomenology is the ability to tune the density of carriers and the strength of their couplings to such modes. To gain control over the electronic density of states, advanced materials have been designed, \textit{e.g.} intercalated graphite \cite{dresselhaus1981intercalation}, doped fullerenes \cite{hebard1991potassium}, and doped charge-transfer insulators \cite{bednorz1986possible}. However, chemical doping often increases the disorder in the system, thus limiting the beneficial effect of modulating the charge density at will. Alternatively, photodoping \cite{yu1991transient} or phonon pumping \cite{mankowsky2016non} have been attempted to control the electronic density of states or the coupling between carriers and collective modes. The advantage of such an approach is that light does not induce structural disorder and can access new states of matter that only exist out of equilibrium.

Ultrafast techniques using visible light pulses have unveiled some of these phenomena by delivering excess energy to the electrons via an intense pump pulse and subsequently monitoring the transfer of such energy to the different underlying bosons via a delayed optical probe. Whenever a preferential electron-boson interaction channel exists, it dominates the carrier thermalization via simultaneous heating of the bosonic mode. In this regard, a body of work has demonstrated the emergence of hot optical phonon effects in semimetals such as graphite \cite{ref:kampfrath, heinz_prb09, ref:breusing} and graphene \cite{heinz_prl10, ref:breusing2011}, where a strong electron-phonon coupling arises due to the ineffective screening of the Coulomb interaction. In this strong coupling regime, the photoexcited nonthermal electron distribution has an increased probability to generate hot optical phonon modes before degrading into a lower energy quasiequilibrium distribution. A similar mechanism was shown to rule the normal-state ultrafast dynamics of cuprate superconductors and proposed to originate from the coupling with high-energy spin fluctuations \cite{ref:dalconte}.

A step further in hot phonon research involves the observation of an anisotropic coupling between different subsets of carriers and a specific phonon. In this respect, a prototypical system is represented by MgB$_2$, which below 39 K is a rare example of two-band phonon-mediated superconductor in the strong coupling regime \cite{ref:nagamatsu}. In its normal state, MgB$_2$ possesses two distinct types of electronic bands crossing the Fermi energy $\mathrm{E_F}$: The quasi-two-dimensional hole-like $\sigma$ bands and the three-dimensional $\pi$ bands. The holes of the $\sigma$-bands, located along the $\Gamma$-A direction of the Brillouin zone, interact very strongly with the branch of the $\mathrm{E_{2g}}$ bond-stretching mode (characterized by an energy of $\sim$ 80 meV at $\Gamma$), displaying a coupling three-times larger than the $\pi$ bands carriers~\cite{ref:kong, yelland, ref:choi, ref:golubov}. So far, the existence of a highly anisotropic electron-phonon coupling in the $\sigma$- and $\pi$-bands has been accessed experimentally only in an indirect way, via quantum oscillations \cite{yelland} or as a signature of the two superconducting (SC) gap values \cite{ref:souma, ref:quilty}. To overcome this limitation, one needs to disentangle the real-time dynamics of both types of carriers, by selectively monitoring suitable observables under nonequilibrium conditions. This aspect is of pivotal interest as it sheds light on a fascinating topic that has remained elusive over the years: The mechanism responsible for connecting the two bands, which also lies at the origin of the sizes and temperature dependence of the SC gaps.

Here, we use an ultrashort 1.55 eV laser pulse to set the $\sigma$ and $\pi$ electronic subsystems of MgB$_2$ out of equilibrium and follow the $\sigma$-$\pi$ interband scattering by probing the variation of the sample reflectivity ($\Delta$R/R) with a broadband pulse covering the in-plane (a-axis) and out-of-plane (c-axis) plasma edges~\cite{ref:guritanu, ref:kakeshita}. In our experiment, the $\sigma$ carrier response manifests itself in a blueshift of the a-axis bare plasma frequency $\omega_{p,a}$ during the first 170 fs after photoexcitation. Such an observation is at odds with the single-band paradigm, in which a heating of the electronic carriers leads to a redshift of $\omega_{p,a}$ \cite{toschi_prl05, kuzmenko_prb07}. We find that this blueshift of $\omega_{p,a}$ can be understood only within the context of a multiband system in which the Fermi surface areas change due to the interaction of the $\sigma$ carriers with strongly-coupled hot phonons. After this initial response, we observe a delayed blueshift of the c-axis plasmon on a longer time scale, which is the signature of weak interband scattering mechanisms allowing for crosstalk between the $\sigma$ and $\pi$ bands. Our work shows that the photoinduced creation of a hot phonon bath in a multiband system can be used to selectively trigger a transient increase of the number of carriers in a given band, opening new perspectives for the selective carrier-density manipulation via near-infrared light. 

In our experiments, we use a high-quality (0001)-oriented MgB$_2$ thin-film, grown by hybrid physical-chemical vapor deposition. In our geometry, the pump beam hits the sample at normal incidence and it is polarized in-plane. The probe beam is a broadband continuum covering the 1.77-2.90 eV spectral region and directed towards the sample under an angle of 15$^{\circ}$. Consequently, it primarily detects the in-plane plasma edge but also explores the c-axis response. Details on the methods are reported in the Supplemental Materials (SM).

\begin{figure}[t]
\begin{center}
\includegraphics[width=\columnwidth]{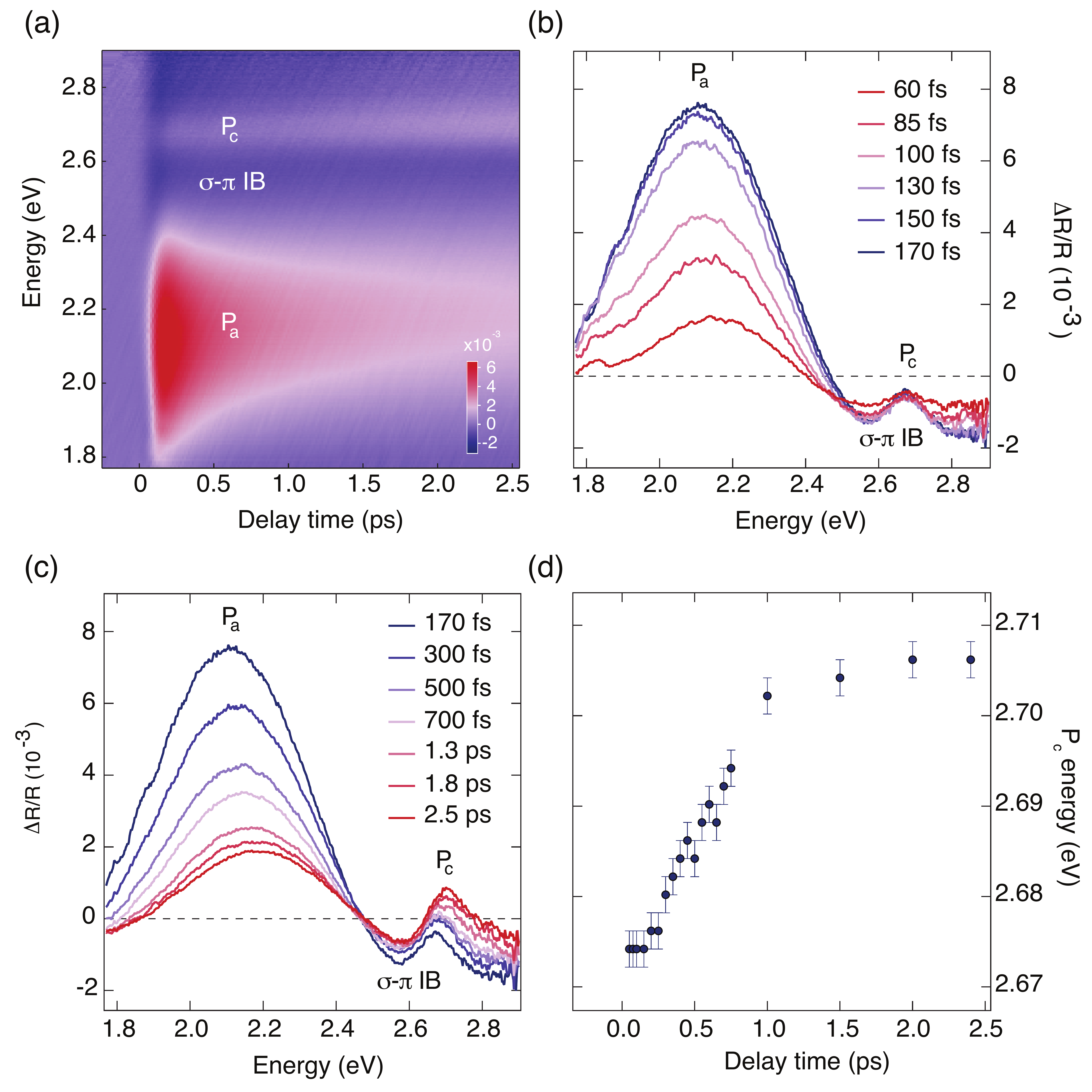}
\caption{(a) Color-coded map of $\Delta$R/R at 10 K as a function of probe photon energy and time delay between pump and probe. The pump photon energy is 1.55 eV and the absorbed fluence is 1.2 mJ/cm$^2$. (b,c) Transient spectra of $\Delta$R/R for different time delays during (b) the rise (c) the decay of the response. (d) Time evolution of the P$_c$ peak energy.}
\label{fig:Fig1}
\end{center}
\end{figure}

Figure 1(a) shows the color-coded map of $\Delta$R/R at 10 K as a function of the probe photon energy and time delay between pump and probe. Direct inspection of the map reveals three features evolving in time: i) a broad positive contribution centered around 2.15 eV (labelled P$_a$), which has a straightforward correspondence with the in-plane plasma edge (see Fig. S1(b)); ii) a broad negative background covering the 2.55-2.90 eV spectral region, which mirrors the $\sigma$-$\pi$ interband transition (IB); iii) a narrow feature around 2.67 eV (P$_c$), which overlaps with the IB and becomes more pronounced with increasing time delay. This structure lies in correspondence with the c-axis plasma edge. All these features become more evident in the transient spectra at selected time delays. Figures 1(b,c) collect the $\Delta$R/R spectra during the rise and the decay of the response, respectively. During the rise (Fig. 1(b)), the feature P$_a$ undergoes a slight redshift, while the barely observable peak of feature P$_c$ does not evolve in energy. In the decay (Fig. 1(c)), both the low-energy zero-crossing point around 1.80 eV and the peak of P$_a$ shift to higher energies as a function of time; moreover, the feature P$_c$ gradually reinforces its weight and undergoes a pronounced blueshift (Fig. 1(d)).

\begin{figure}[t]
\begin{center}
\includegraphics[width=\columnwidth]{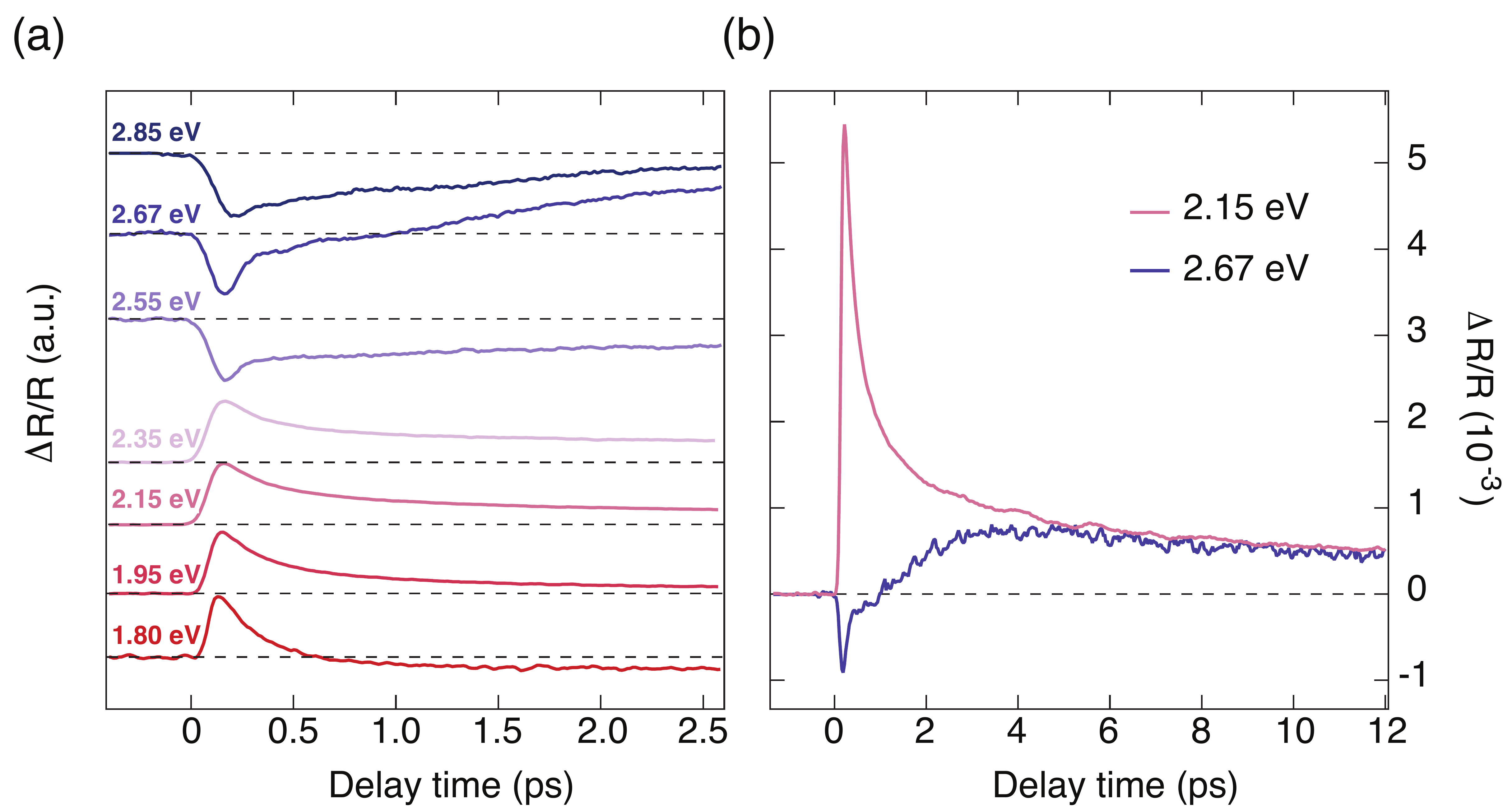}
\caption{(a) Normalized $\Delta$R/R temporal traces up to 2.5 ps at 10 K for different probe photon energies, indicated in the labels. (b) $\Delta$R/R temporal traces up to 12 ps in correspondence of the P$_a$ and P$_c$ spectral features.}
\label{fig:Fig2}
\end{center}
\end{figure}

\begin{figure*}[t]
\begin{center}
\includegraphics[width=2\columnwidth]{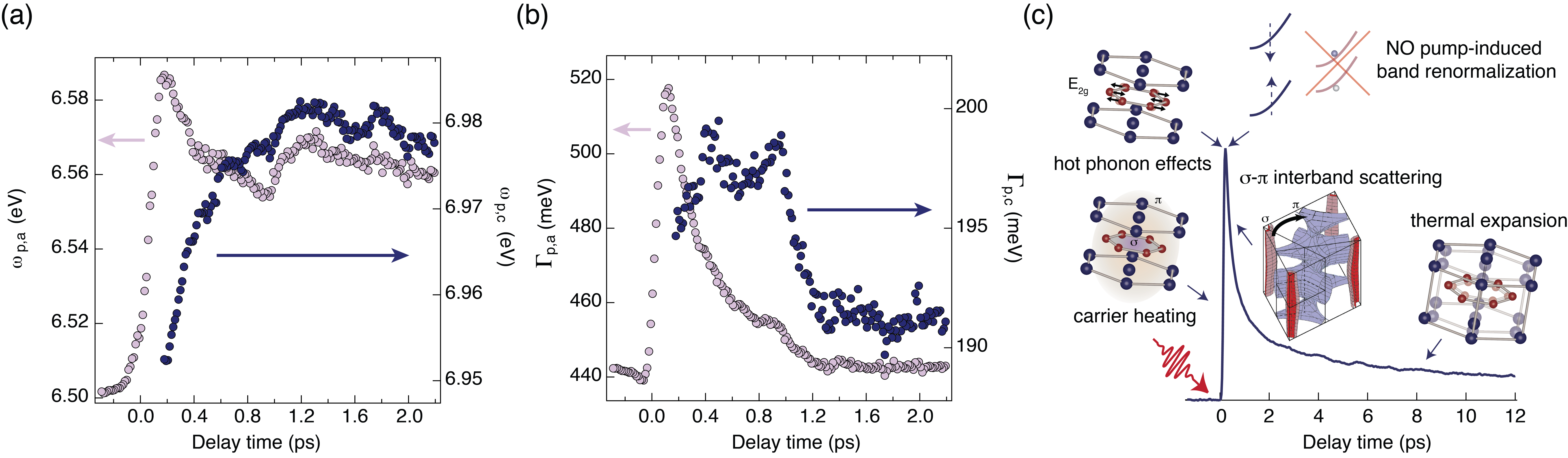}
\caption{Temporal evolution of (a) the bare plasma frequency $\omega_{p,a/c}$ and (b) the optical scattering rate $\Gamma_{p,a/c}$ along the a- (pink dots) and c-axis (blue dots). (c) Cartoon of the ultrafast dynamics. After the interaction with the solid, the ultrashort laser pulse leads to the excitation of both $\sigma$ and $\pi$ carriers. The nonthermal $\sigma$ carriers are strongly coupled to the branch of the $\mathrm{E_{2g}}$ phonon mode and efficiently generate hot phonons during the first 170 fs. Subsequently the energy stored in the hot-phonon subsystem is released to the $\pi$ carriers via interband scattering and to low-energy phonons via anharmonic decay.}
\label{fig:Fig3}
\end{center}
\end{figure*}

Complementary information is provided by the $\Delta$R/R temporal dynamics. Several time traces are reported in Figs. 2(a,b) at representative probe photon energies. We observe that the signal displays a much longer rise time than the response function of our set-up (50 fs), reaching its maximum absolute amplitude at 170 fs. Moreover, the $\Delta$R/R decay depends on the probe photon energy, as the relaxation dynamics in correspondence to the feature P$_c$ (2.67 eV) differs from the one governing the rest of the spectrum. As expected from Fig. 1(c), on the low-energy side (red trace at 1.80 eV in Fig. 2(a)) the signal undergoes a sign change over time, consistent with previous single-wavelength pump-probe experiments in the near-infrared and in the normal state~\cite{ref:xu, ref:demsar}. Our results are rather independent of the sample temperature, except that the feature $\mathrm{P_c}$ becomes sharper for decreasing temperature. The insensitivity of the transient response to temperature is due the absorbed fluence regime (0.2 to 3 mJ/cm$^2$) of our experiment, which exceeds the threshold for the complete vaporization of the SC condensate by two orders of magnitude~\cite{ref:xu, ref:demsar}. At 10 K, we can safely assume that the pump pulse suddenly melts the SC order parameter and drives MgB$_2$ into its normal phase. The recombination dynamics of the quasiparticles into Cooper pairs occurs within hundreds of ps~\cite{ref:xu, ref:demsar}, \textit{i.e.} outside our temporal window.

We first discuss the narrow feature P$_c$. It is characterized by distinct dynamics compared to the remaining visible spectral range, which is indicative of a different spectroscopic origin. A simple phenomenological explanation for the presence of a sharp peak in the $\Delta$R/R spectrum of a metal can be given by recalling that at high frequencies $\epsilon_1$($\omega$)= $\epsilon_{\infty}$(1 - $\omega_p^{*2}$/$\omega^2$), where $\epsilon_{\infty}$ is the high-energy dielectric function, $\omega_p^*$ = $\sqrt{4\pi\widetilde{n}e^2/m\epsilon_{\infty}}$ is the dressed plasma frequency, $\widetilde{n}$ is the carrier density and $m$ is the free electron mass. By assuming that the main effect of the pump pulse is to change $\widetilde{n}$, it results that $\Delta$R/R $\approx$ 8$\pi$e$^2\Delta\widetilde{n}$/m$\omega^2\sqrt{\epsilon_1}$. Hence, $\Delta$R/R displays a peak at $\omega_p^*$, \textit{i.e.} the energy at which $\epsilon_1$ = 0. Therefore, the increase of $\widetilde{n}$ implies a rigid shift of the plasma edge towards higher energies. Providing an explanation to this process is not obvious, since the total carrier density in the system is constant and can just be redistributed energetically by the action of the pump pulse. The microscopic origin behind such an effect will be clarified later. Up to now, following the above argument, we can confirm that P$_c$ corresponds to the c-axis longitudinal plasmon mode, which consistently lies at $\sim$ 2.66 eV in the equilibrium optical spectra~\cite{ref:balassis, ref:guritanu, ref:cai}. The plasmon energy is determined by the IBs from the B $\pi$ and Mg $\zeta$ bands, which possess an almost parallel dispersion in the $\Gamma$-K-M plane of the Brillouin zone~\cite{ref:ku}. Correspondingly, its real-space charge distribution involves coherent charge fluctuations between parallel B and Mg sheets.

\noindent We now focus on the evolution of features P$_a$ and IB. As recent pump-probe experiments covering the 0.75-1.40 eV spectral range of MgB$_2$ have revealed a negative sign in the $\Delta$R/R response~\cite{ref:dalconte}, the positive sign of P$_a$ in our spectra is compatible with an instantaneous broadening of the a-axis plasma edge in the rise of the response. In the relaxation, the a-axis plasma edge narrows and redshifts over time, which is evidenced by the $\Delta$R/R zero-crossing evolution. The negative background in the 2.50-2.85 eV region suggests that an ultrafast change of the $\sigma$-$\pi$ IB has also occurred after the photoexcitation, involving a broadening of its shape.

To disentangle the carrier dynamics along the a- and c-axis and extract the significant parameters of the response, we fit an anisotropic Drude-Lorentz model to the static and time-resolved data at 300 K (see SM). By iterating this procedure over the probed time window, we can map the temporal evolution of the bare plasma frequencies ($\omega_{p,a/c}$ in Fig. 3(a)) and optical scattering rates ($\Gamma_{p,a/c}$ in Fig. 3(b)), which respectively reflect the position and damping of the plasma edge. We obtain an ultrafast blueshift of the a-axis plasma frequency $\omega_{p,a}$ (violet dots, Fig. 3(a)), accompanied by a simultaneous increase of the a-axis scattering rate $\Gamma_a$ (violet dots, Fig. 3(b)). After 170 fs, $\omega_{p,a}$ redshifts on a timescale of several ps, while $\Gamma_a$ relaxes to its equilibrium value within $\sim$ 1 ps. While the dynamics of $\Gamma_a$ agrees with the results of Ref. \cite{ref:dalconte}, the temporal evolution of $\omega_{p,a}$ is a completely new observable that becomes apparent only when the broadband probe is tuned to our spectral region. Signatures of this behavior were already evident in Ref. \cite{ref:demsar}, but never discussed. More remarkable is the temporal evolution of the c-axis Drude contribution (blue dots, Figs. 3(a,b)): $\omega_{p,c}$ undergoes a slow blueshift within 1 ps, followed by its stabilization around a constant value; instead, $\Gamma_c$ slowly rises on a timescale of $\sim$ 300 fs, before decreasing again after 1 ps.

To correctly interpret the intriguing dynamics of $\omega_{p,a/c}$, we pursue theoretical/computational calculations through three distinct temporal regimes. To unravel the effect of the pump pulse on the system, we perform a frozen-phonon \textit{ab initio} calculation of the coherent electron dynamics triggered by an ultrashort laser pulse (see SM). The aim is to clarify whether carrier density-dependent many-body effects can renormalize the band structure and lead to the observed behavior. Remarkably, we can explain neither the initial blueshift of $\omega_{p,a}$ nor the subsequent opposite trends of $\omega_{p,a}$ and $\omega_{p,c}$. This calls for a deeper investigation of dynamical many-body effects involved in such an anomalous phenomenon. 

For this reason, we perform analytical calculations to demonstrate the role played by hot phonons in the unconventional blueshift of $\omega_{p,a}$ within the initial 170 fs. A cartoon summarizing the various steps of the ultrafast dynamics is shown in Fig. 3(c). At equilibrium, many-body effects manifest themselves in the strong coupling between the electrons and the lattice degrees of freedom. In particular, the $\sigma$ bands are strongly coupled (coupling constant $\lambda \approx 1$) with the phonon branch of the $\mathrm{E_{2g}}$ mode, mainly at small momenta; a residual electron-phonon interaction ($\lambda \approx 0.3$ in average) connects the $\sigma$- and $\pi$-bands. Within this multiband scenario at T = 0, a band $\beta$ induces a shift of $\mathrm{E_F}$ on a band $\alpha$ due to the strong coupling with a phonon of energy $\hbar\omega_{\rm ph}$. This shift can be quantified as $\chi_\alpha \propto \lambda_{\alpha\beta} \ln(W^{\rm top}_\beta/W^{\rm bottom}_\beta)$, where $W^{\rm top}_\beta$, $W^{\rm bottom}_\beta$ are the bandwidths of the band $\beta$ above and below $\mathrm{E_F}$, respectively. For low-density bands (as in MgB$_2$ or in pnictides), within the parabolic-band approximation, such a shift of $\mathrm{E_F}$ induces a change of the charge density $\Delta \widetilde{n}_\alpha \approx m \chi_\alpha/\pi$. These effects have been observed in pnictides in de Haas-van Alphen \cite{coldea} and angle-resolved photoemission experiments \cite{ding2011electronic}. During the photoexcitation, the pump provides energy to the carriers in both bands, bringing the $\sigma$ and $\pi$ carriers to high temperatures $\mathrm{T_{e,\sigma}}$ and $\mathrm{T_{e,\pi}}$. While in conventional metals the phonon bath remains at $T_{\rm ph} \ll T_{\rm e}$, in MgB$_2$ the scenario is expected to differ dramatically. Indeed, the $\sigma$ holes are strongly coupled to the single $\mathrm{E_{2g}}$ phonon mode. Thus, this phonon also becomes hot after the photoexcitation. As a consequence, we expect that both $\mathrm{T_{e,\sigma}}$ and $\mathrm{T_{ph}}$ increase right after the interaction with the pump pulse. In contrast, as the $\pi$ carriers are not strongly coupled to this mode, only $\mathrm{T_{e,\pi}}$ is expected to increase. As a result, it is necessary to elucidate the effect produced on $\omega_{p,a/c}$ by an increase of the electronic and bosonic temperatures. As shown in the SM, the heating of the $\mathrm{E_{2g}}$ phonon to a high temperature ($\mathrm{T_{ph}}$ $\approx$ 450 K) has a novel and dominant effect on the effective charge density of the in-plane $\sigma$ bands, explaining quantitatively the remarkable initial increase of $\omega_{p,a}$. The electron-phonon energy exchange is completed within 170 fs and the subsequent decay of the a-axis signal is governed by the release of the energy stored in the hot $\mathrm{E_{2g}}$ phonon subsystem to the remaining electronic degrees of freedom via phonon-electron and electron-electron scattering. At this stage, interband scattering with the $\pi$ electrons leads to an increase of the charge carriers in the $\pi$ band, needed to restore the overall charge density of the system, arising from a balance between hole-like and electron-like bands. This delayed response reflects in the increase of $\omega_{p,c}$, controlled solely by the $\pi$ carrier subset \cite{ref:guritanu, ref:mazin}, and occurs on the $\sim$ 2 ps timescale of the weaker $\sigma$-$\pi$ scattering processes.

Once the electronic and hot phonon subsystems have thermalized, the subsequent release of energy occurs via anharmonic decay to acoustic phonons, lasting tens of ps \cite{ref:demsar,heinz_prb09}. We also model such long timescale dynamics by calculating how the c-axis plasmon energy is modified for an increased lattice expansion (see SM). We find a sizeable redshift of the $c$-axis plasmon energy, which is also consistent with our experimental results.

In conclusion, we combined ultrafast optical spectroscopy and state-of-the-art calculations to map the nonequilibrium dynamics in MgB$_2$ and reveal the phonon-mediated $\sigma$-$\pi$ interband scattering process. Our results show that ultrafast light excitation can be used to control the carrier density in multiband materials via hot phonon effects. More generally, by setting a multiband system out-of-equilibrium, the microscopic details of the scattering mechanism between distinct electronic subsets can be unraveled via selective observables in the time domain.

\begin{acknowledgments}
Work at LUMES was supported by NCCR MUST. V.M.S acknowledges the partial support from the University of the Basque Country UPV/EHU, Grant No. IT-756-13, and the Spanish Ministry of Economy and Competitiveness MINECO, Grant No. FIS2013-48286-C2-1-P. Work at Temple University was supported by the U.S. Department of Energy, Office of Science, High Energy Physics, under Award No. DE-SC0011616. L.B. acknowledges financial support by Italian MIUR under projects FIRB-HybridNanoDev-RBFR1236VV, PRIN-RIDEIRON-2012X3YFZ2, and Premiali-2012  AB-NANOTECH.
\end{acknowledgments}

\clearpage
\newpage

\setcounter{section}{0}
\setcounter{figure}{0}
\renewcommand{\thesection}{S\arabic{section}}  
\renewcommand{\thetable}{S\arabic{table}}  
\renewcommand{\thefigure}{S\arabic{figure}} 
\renewcommand\Im{\operatorname{\mathfrak{Im}}}
\titleformat{\section}[block]{\bfseries}{\thesection.}{1em}{} 

\section{S1. Sample preparation and mounting}

The MgB$_2$ films were grown by hybrid physical-chemical vapor deposition at a substrate temperature of 730 $^{\circ}$C, hydrogen carrier gas pressure of 40 Torr, hydrogen flow rate of 400 sccm, and diborane  mixture (5$\%$ B$_2$H$_6$ in H$_2$) flow rate of 20 sccm. The deposition rate was 55 nm/min. The resulting thin film was a parallelepiped with a (0001) surface of 5 mm $\times$ 5 mm and a thickness of 200 nm along the c-axis. This thickness was controlled by the deposition time (220 s). Epitaxial MgB$_2$ films were deposited directly on SiC-(0001) substrates. After the deposition, the sides of the substrate were cleaned with 10$\%$ hydrogen chloride acid to remove the MgB$_2$ deposit. The structure of the films was characterized by x-ray diffraction and cross-sectional transmission electron microscopy using a JEOL 2100 LaB6 operated at 200 kV. The dc transport properties were measured using the four-probe van der Pauw method.

It is well established that surface contamination by exposure to air strongly reduces the absolute reflectivity of MgB$_2$, decreasing the value of $\omega_p$ over time \cite{ref:guritanu, ref:kakeshita, ref:kuzmenko}. To prevent the measurements to suffer from this effect, the sample was stored in a vacuum environment before being introduced in the cryostat for the ultrafast optical measurements. During the transfer from the sample container to the cryostat, the thin film was kept in an Argon flow until vacuum was produced in the cryostat. During the measurements, the achieved vacuum pressure was $\sim$ 10$^{-8}$ mbar at room temperature and $\sim$ 10$^{-9}$ mbar at 10 K. This vacuum pressure was sufficient to maintain the sample properties unaltered for more than one month, allowing for extensive measurements at different temperatures. For the low-temperature measurements, to counteract the adsorption of carbon-oxide compounds on the MgB$_2$ thin film surface, the sample was warmed up above $\sim$ 230 K after about 2 hours at low temperatures. This treatment restored the reflectivity of its surface.

\section{S2. Experimental setup}

Femtosecond broadband transient reflectivity experiments have been performed using a set-up described in Ref. \cite{baldini2016versatile}. A Ti:Sapphire oscillator, pumped by a continuous-wave Nd:YVO$_4$ laser, emitted sub-50 fs pulses at 1.55 eV with a repetition rate of 80 MHz. The output of the oscillator seeded a cryo-cooled Ti:Sapphire amplifier, which was pumped by a Q-switched Nd:YAG laser. This laser system provided $\sim$ 45 fs pulses at 1.55 eV with a repetition rate of 6 kHz. One third of the output, representing the probe beam, was sent to a motorized delay line to set a controlled delay between pump and probe. The 1.55 eV beam was focused on a 3 mm-thick CaF$_2$ cell using a combination of a lens with short focal distance and an iris to limit the numerical aperture of the incoming beam. The generated continuum covered the 1.77-2.90 eV spectral range. The probe was subsequently collimated and focused onto the sample through a pair of parabolic mirrors under an angle of 15$^\circ$. The remaining two thirds of the amplifier output, representing the pump beam, were directed towards the sample under normal incidence. Along the pump path, a chopper with a 60 slot plate was inserted, operating at 1.5 kHz and phase-locked to the laser system. Both pump and probe were focused onto the sample with spatial dimensions of \mbox{120 $\mathrm{\mu m}$ $ \times$ 87 $\mathrm{\mu m}$} for the pump and 23 $\mathrm{\mu m}$ $\times$ 23 $\mathrm{\mu m}$ for the probe. The sample was mounted inside a closed cycle cryostat, which provided a temperature-controlled environment in the range 10-340 K. The reflected probe was dispersed by a fiber-coupled 0.3 m spectrograph and detected on a shot-to-shot basis with a complementary metal-oxide-semiconductor linear array.

\section{S3. Anisotropic differential Drude-Lorentz model}

\begin{figure}[b]
\begin{center}
\includegraphics[width=\columnwidth]{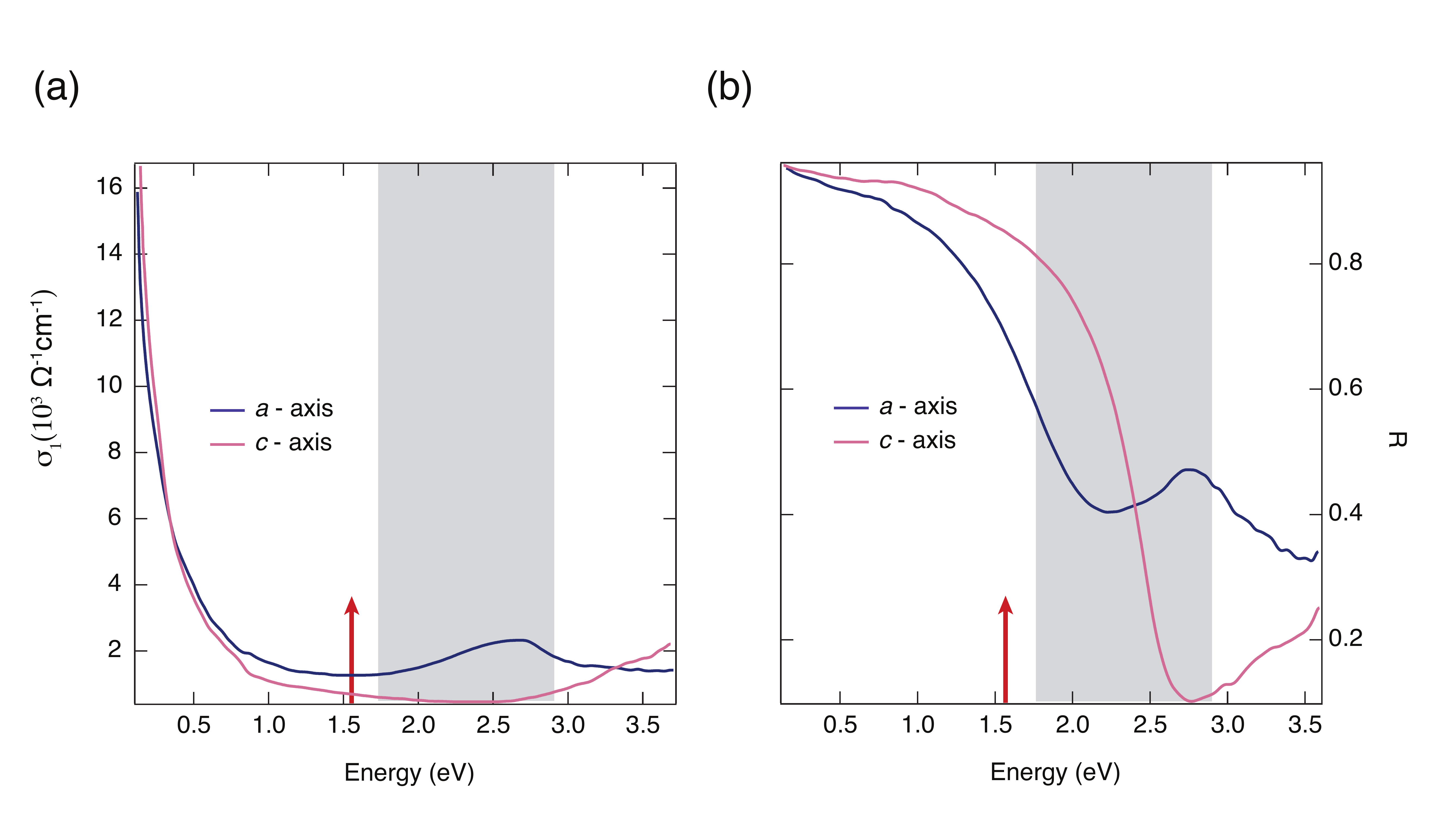}
\caption{(a) Real part of the optical conductivity and (b) reflectivity of MgB$_2$ measured via SE at RT along the a- (blue curves) and c-axis (pink curves). The data have been adapted from Ref. \cite{ref:guritanu}. The red arrow indicates the pump photon energy and the grey shaded area the probing range of the nonequilibrium experiment.}
\label{fig:Static}
\end{center}
\end{figure}

\begin{figure}[t]
\begin{center}
\includegraphics[width=\columnwidth]{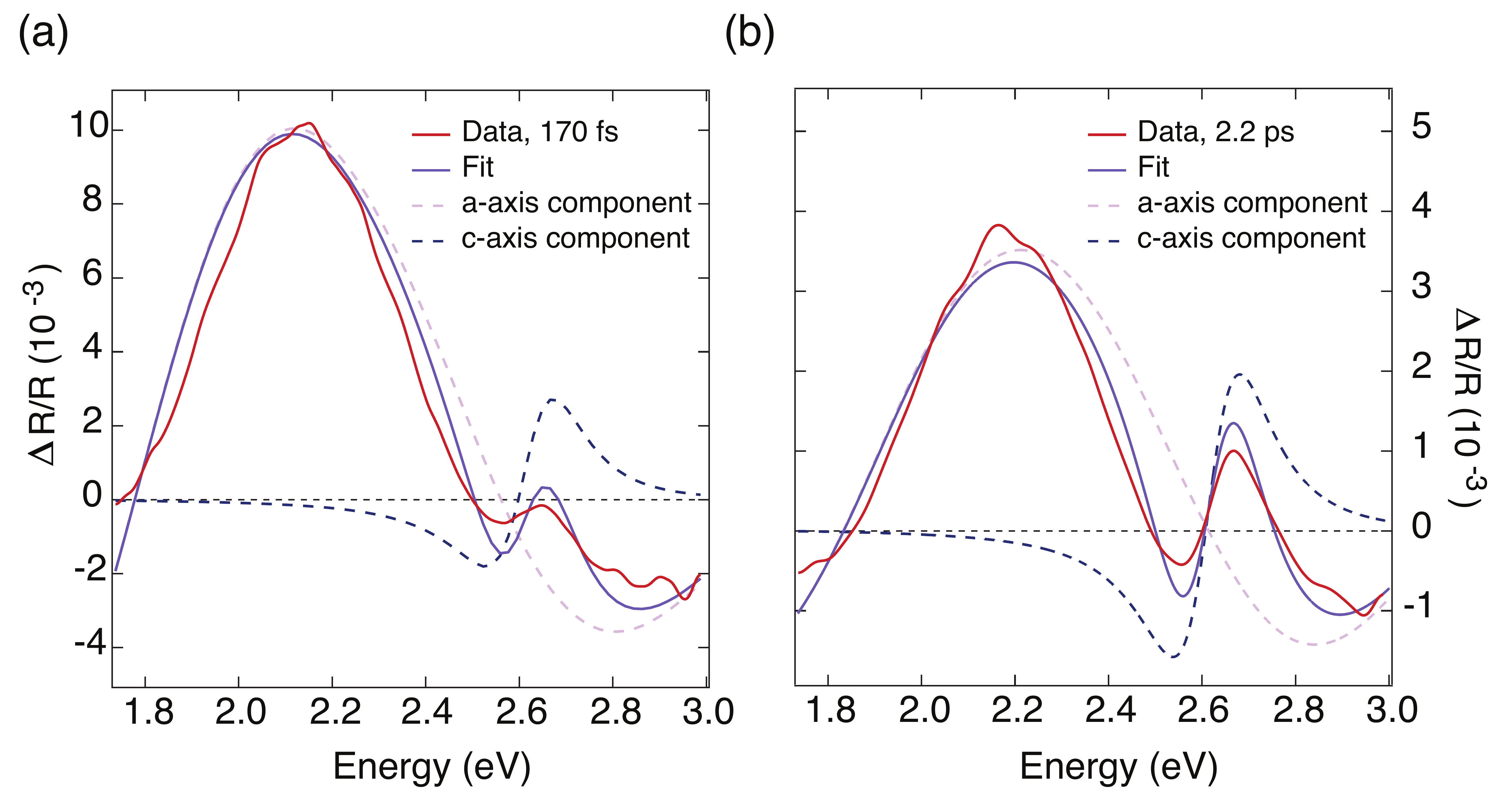}
\caption{Results of the anisotropic differential Drude-Lorentz model on the $\Delta$R/R spectrum at 300 K for a delay time of (a) 170 fs and (b) 2.2 ps. The original data are represented as solid red lines, the resulting fit as solid violet lines, the a- and c-axis contributions in dashed pink and blue lines, respectively.}
\label{fig:FigS1}
\end{center}
\end{figure}

To disentangle the a- and c-axis contributions to the $\Delta$R/R caused by the 15$^\circ$ angle of incidence of our probe beam, we fit an anisotropic Drude-Lorentz model to the static data at 300 K displayed in Fig. \ref{fig:Static}(a,b) (adapted from Ref. \cite{ref:guritanu}). Due to the higher quality of our sample compared to the one used in Ref. \cite{ref:guritanu}, we slightly increase the c-axis bare plasma frequency value to achieve a better fit. The spectral features covered by our probe range are the a-axis plasma edge around 2.00 eV, the a-axis $\sigma$-$\pi$ interband transition around 2.60 eV and the c-axis plasma edge around 2.62 eV (grey shaded areas in Fig. \ref{fig:Static}(a,b)). We can therefore limit the fit of our model to six free parameters: The bare plasma frequencies $\omega_{p,a}$ and $\omega_{p,c}$ along the a- and c-axis, the corresponding dampings $\Gamma_{a}$ and $\Gamma_{c}$, and the plasma frequency and optical scattering rates of the a-axis $\sigma$-$\pi$ interband transition. This provides a reliable fit of the $\Delta$R/R data while enabling us to identify their physical origin. The fit is performed separately for the spectra at 300 K (to match the temperature of the steady-state data) at each time delay, covering a temporal window up to 2.4 ps. The robustness of the fit is evidenced by Fig. \ref{fig:FigS1}(a,b), where we show the results in correspondence to the rise (170 fs) and decay (2.2 ps) of the response. We observe that the fit matches the experimental data with good precision and yields the contribution from the a- and c-axis response. We underline that the fit does not converge if $\omega_{p,a/c}$ are not left free but are forced to redshift after the interaction with the pump pulse. This also confirms the reliability of the results presented in the main text.

\section{S4. Excitation mechanism of the pump pulse}
\label{Excitation_MgB2}

To correctly interpret the observed ultrafast dynamics, it is pivotal to clarify how the electronic subsystem of MgB$_2$ is affected by the interaction with the near-infrared pump pulse at 1.55 eV. Since this pump photon energy lies below 2.40 eV, it is not sufficient to efficiently induce the in-plane $\sigma$-$\pi$ interband transition. On one hand, we can expect that the tails of both oscillators associated with the $\sigma$-$\sigma$ and the $\sigma$-$\pi$ interband transitions can provide a contribution to the excitation process \cite{ref:guritanu}. On the other hand, energy transfer from the laser to the electron gas is expected to occur mainly via free-carrier absorption processes. This idea is confirmed by observing that the laser pulse at 1.55 eV lies also on the tail of the Drude response in the a-axis optical conductivity shown in Fig. \ref{fig:Static}(a). However, it is well established that the a-axis optical conductivity of MgB$_2$ can be described only by assuming a two-component Drude response, due to the contributions of both $\sigma$ and $\pi$ carriers \cite{ref:guritanu, ref:kakeshita}. Thus, it becomes crucial to establish whether our laser pulse at 1.55 eV selectively excites one type of carriers or if it exchanges energy with both electronic subsystems.  

\begin{figure}[t]
\begin{center}
\includegraphics[width=\columnwidth]{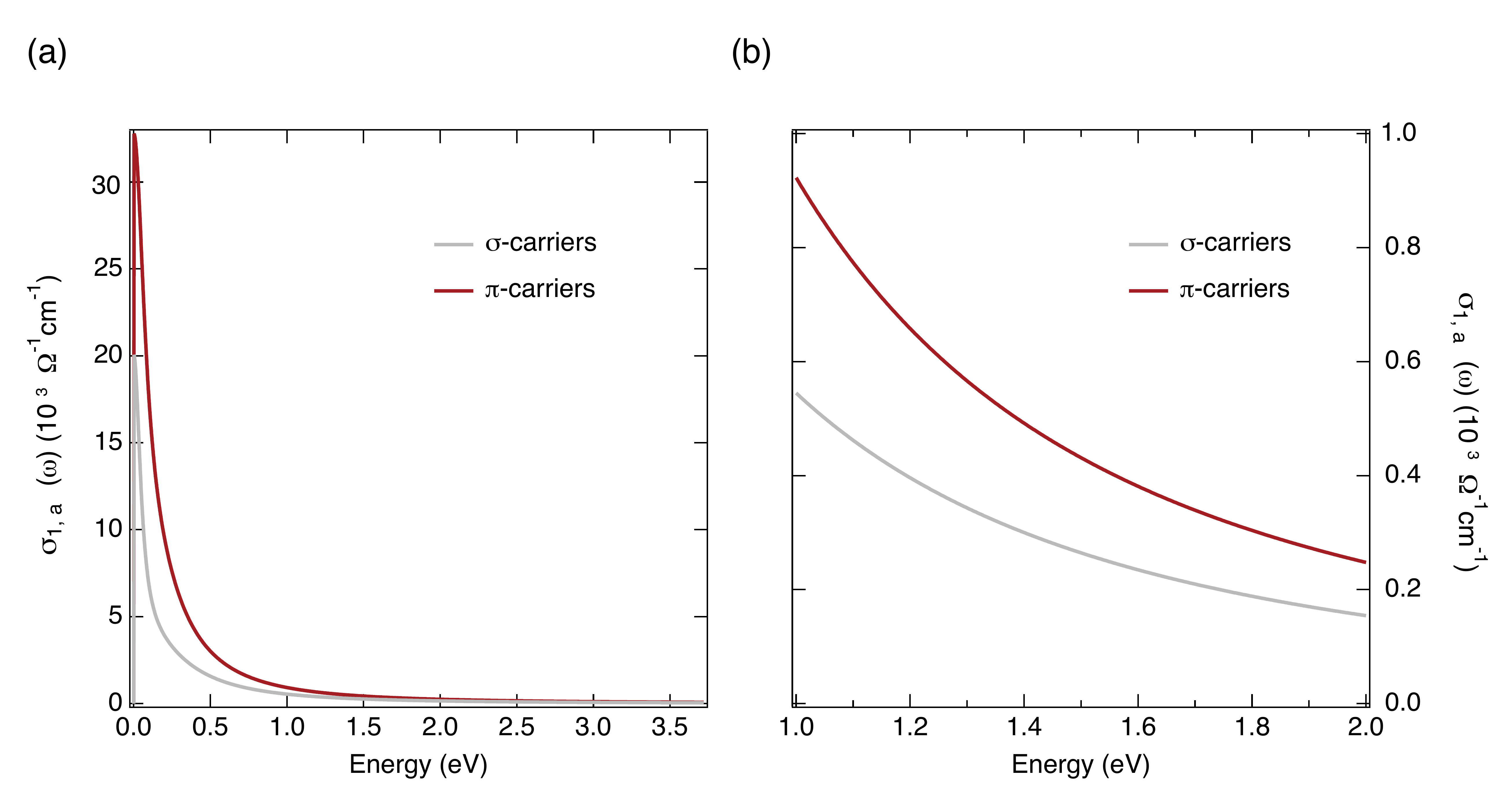}
\caption{(a) Relative contribution of the $\sigma$ and $\pi$ carriers to the equilibrium Drude response, on the basis of the calculations reported in Ref. \cite{ref:guritanu}. (b) Zoom of the calculations reported in (a) in the vicinity of the pump photon energy at 1.55 eV.}
\label{fig:Contributions}
\end{center}
\end{figure}

To this end, we quantify the relative contribution of the $\sigma$ and $\pi$ carriers to the equilibrium Drude response at 1.55 eV by performing an \textit{ab initio} simulation of the optical properties of MgB$_2$ and including one type of electronic subsystem at a time. The two contributions to the optical conductivity are evaluated on the basis of the calculations reported in Ref. \cite{ref:guritanu}. The results are shown in Fig. \ref{fig:Contributions}(a,b) and show that both electronic subsystems provide a sizeable response to the incoming electromagnetic field at 1.55 eV. In particular, the Drude contribution of the $\pi$ carriers is found to be slightly more pronounced at the laser photon energy.

As a result, we can conclude that our pump pulse is not able to selectively excite the carriers residing in either the $\sigma$ or the $\pi$ band and that both electronic subsystems undergo an increase of their temperature $\mathrm{T_{e,\sigma}}$ and $\mathrm{T_{e,\pi}}$ upon interaction with the pump pulse. Irrespective of the details of the photoexcited carrier density, it is important to observe that our pump pulse lies in the near-infrared spectral range, thus enabling the photoexcited carriers to generate high-energy phonons.

\section{S5. Photoinduced renormalization of the electronic structure}

To correctly interpret the blueshift displayed by the a- and c-axis plasma edges, we first need to clarify how the electronic subsystem of MgB$_2$ is affected by the interaction with the near-infrared pump pulse at 1.55 eV. To verify this hypothesis, we need to discard the possibility that many-body effects caused by electron-electron interactions play an important role in the renormalization of the electronic structure. Hence, we study how the electronic structure of MgB$_2$ is modified after being photoexcited by an ultrashort laser pulse. 

Explicit time-dependent quantum mechanical calculations can provide information on the coherent electron dynamics generated by intense short laser pulses and give a microscopic description of the fast electronic response in photoexcited systems with attosecond resolution. The in-house developed code implements a quantum-mechanical time-dependent method that evolves the electronic wavefunction with the propagator proposed by Allen and co-workers~\cite{ref:allen, ref:graves, ref:torralva, ref:dou}. The evolution of the electronic wavefunction is therefore calculated with a generalized Cayley algorithm, based on a Dyson-like expansion of the time-evolution operator~\cite{ref:allen2}. Numerically, the time-dependent Schr\"odinger equation (TDSE) is solved by the Cranck-Nicolson approximation method~\cite{ref:crank, ref:zgsun}, based on expressing the exponential operator e$^{H\Delta t}$ with second-order accuracy. It uses the commonly recommended finite differences scheme to approximate the solutions of the TDSE in complex multiscale phenomena. This is a unitary and stable approach, which conserves probability and preserves orthogonality; moreover, as an O(N) method, it is particularly appropriate to investigate large systems, such as complex materials and biological molecules.

\begin{figure}[b]
\begin{center}
\includegraphics[width=\columnwidth]{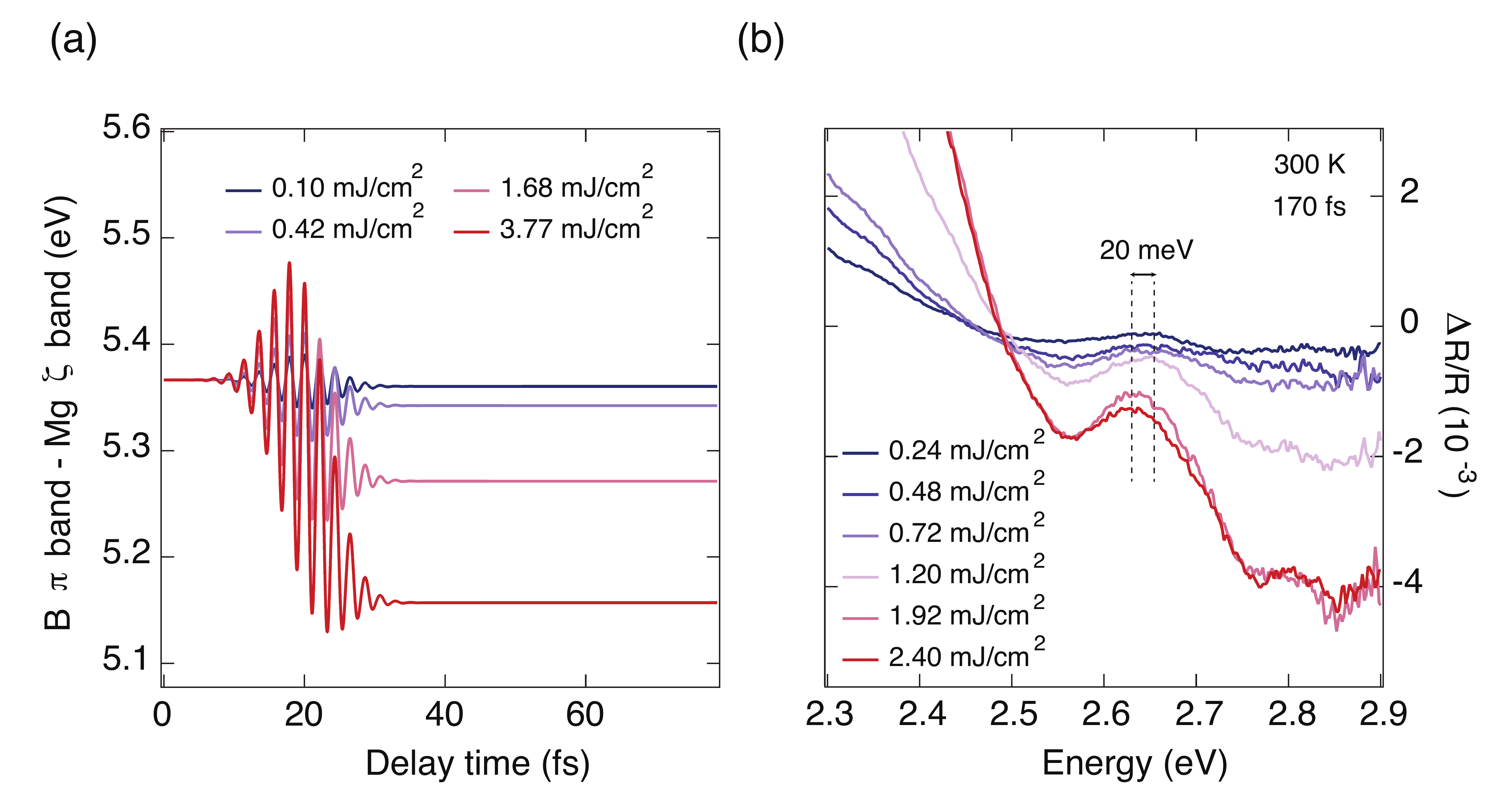}
\caption{(a) Calculated ultrafast renormalization of the B $\pi$ - Mg $\zeta$ interband transition energy at the $\Gamma$ point of the BZ. The results are reported for different field strengths (indicated in the label), which are comparable to the experimental conditions. (b) Transient spectra of $\Delta$R/R in the vicinity of the c-axis plasmon peak at 300 K for different absorbed fluences and at a delay time of 170 fs.}
\label{fig:FigS3}
\end{center}
\end{figure}

\noindent The propagator contains the sum of the unperturbed Hamiltonian at a certain k point of the crystal periodic structure (H$^0$(k)) and of the perturbing Hamitonian (H$^{\prime}$), representing the interaction of the applied electromagnetic field with the crystal dipole moment matrix of the periodic system. It reads
\begin{equation}
\Psi_j (t+\Delta t, k) = \Big( 1+\frac{iH(k) \Delta t}{2\hbar} \Big) ^{-1} \Big (1- \frac{iH(k) \Delta t}{2\hbar}\Big) \Psi_j (t, k)
\end{equation}
\noindent The initial electronic wavefunction, $\Psi$(t$_0$, k), is the matrix of crystal orbitals (CO's) of the periodic system at a specific k point, built as a linear combination of Bloch orbitals (BO's) extended overall the entire crystalline solid; the BO's are, in turn, a linear combination of a set of basis functions; $\Delta$t is the time step; H(k) is the Hamiltonian of a specific k point in the reciprocal space, H(k) = H$^0$(k)+ H$^\prime$. The perturbing Hamiltonian is then calculated as a product of the electric field and the crystal dipole moment, here written as
\begin{equation}
H(t) = -D~E(t)~\sin(\omega t)~e^{-t/\tau}
\end{equation}
\noindent where, D is the dipole moment matrix between crystal orbitals, $\omega$ is the frequency of the incoming photon; E(t) is the electric field amplitude or field envelope, here kept constant in time, and $\tau$ is the pulse duration, here set at 6 fs. The calculations are carried out in the CO framework. The field polarization can be chosen during simulations. The CO's coefficients of $\Psi$(t, k) are updated at each time step under the effect of the external field. The propagation is carried out starting from COs orthonormalized by the L\"owdin transformation. 

\noindent This approach allows to gain a detailed microscopic description of the ultrafast electronic response of the system, which can be followed in terms of a large number of observables (\textit{i.e.} time-dependent electron density distribution, CO occupancies, dipole moment changes and relative absorption spectra). In the past, it has been successfully employed also to investigate the photolytic splitting of water ice by vacuum ultraviolet light~\cite{ref:acocella1}, the dynamics of photo-activated chemical bonding~\cite{ref:acocella2}, electron transfer processes~\cite{ref:acocella3}, photoinduced dissociation~\cite{ref:jones1}, nonlinear optical properties~\cite{ref:jones2}, mono- and multi-photon excitations~\cite{ref:acocella4}, and photoelectron spectra of coronene~\cite{ref:acocella5}. Using this technique, we investigate the ultrafast changes of the MgB$_2$ band structure during the first 80 fs after photoexcitation at selected critical momentum points of the Brillouin zone (BZ).

The ground electronic states of the structures were calculated with the Gaussian 09~\cite{ref:gaussian09} suite of programs with periodic boundary conditions. The nuclear geometries are frozen during the coherent electron dynamics so that the pure electron reorganization of the system wavefunction is investigated. The MgB$_2$ calculations are run on the hexagonal AlB$_2$-type lattice structure~\cite{ref:jones3}, with space group P6/mmm and lattice parameters, a = 3.084 \AA~ and c = 3.522 \AA, using the Perdew, Burke and Ernzerhof functional~\cite{ref:perdew} implemented in Gaussian09. A 724 k-points grid in the reciprocal space is created. The basis sets applied are the Small Split-Valence 3-21SP function for Mg atoms and the 6-31G for B atoms.

The effect of a single pump laser beam, oriented in the plane of the structure layers, is investigated on the fast time scale at the $\Gamma$, A, K and M points of the BZ. The field-induced changes of the electronic structure are calculated using a pump photon energy at 1.55 eV or tuned to resonate with the selected k points. Moreover, in order to investigate the electronic response dependence on the pump fluence, different field strengths are applied during the dynamics, corresponding to fluences of 0.10, 0.42, 1.68, 3.77, 6.72 and 10.51 mJ/cm$^2$. Tables 1 and 2 collect, respectively, the net single-particle energy differences calculated as the difference between initial (at 0 time) and final (after 40 fs) highly-occupied CO (HOCO) - lowest-unoccupied CO (LUCO) gap for $\Gamma$, A, M and K points for a pump photon energy at 1.55 eV or in resonance with the different k points. Notice that a negative value in the tables indicates a decrease in the single-particle band energies, while a positive value corresponds to their increase. 

\begin{table}[h]
\centering
\label{TableI}
\begin{tabular}{|c|c|c|c|c|}
\hline
\begin{tabular}[c]{@{}c@{}}Fluence\\ (mJ/cm$^2$)\end{tabular} & \begin{tabular}[c]{@{}c@{}}$\Gamma$\\ (eV)\end{tabular} & \begin{tabular}[c]{@{}c@{}}A\\ (eV) \end{tabular} & \begin{tabular}[c]{@{}c@{}}M\\ (eV) \end{tabular} & \begin{tabular}[c]{@{}c@{}}K\\ (eV) \end{tabular} \\ \hline
0.10                                                       & -0.009 & 0  & 0  &  0 \\ \hline
0.42                                                       &  -0.036                    &  0 & 0  &  0 \\ \hline
1.68                                                       &          -0.141            & 0  &  0 &  0 \\ \hline
3.77                                                       &   -0.310                   &  0 & 0  & 0  \\ \hline
6.72                                                       &   -0.532                   &  0 &  0 & 0  \\ \hline
10.51                                                      &   -0.794                   & 0  &  0 &  0 \\ \hline
\end{tabular}
\caption{Net band energies renormalization calculated upon excitation with a 1.55 eV pump pulse at different points ($\Gamma$, A, M, K) of the BZ and for different absorbed pump fluences.}
\end{table}

\begin{table}[h]
\centering
\label{TableII}
\begin{tabular}{|c|c|c|c|c|}
\hline
\begin{tabular}[c]{@{}c@{}}Fluence\\ (mJ/cm$^2$)\end{tabular} & \begin{tabular}[c]{@{}c@{}}$\Gamma$\\ (eV)\end{tabular} & \begin{tabular}[c]{@{}c@{}}A\\ (eV) \end{tabular} & \begin{tabular}[c]{@{}c@{}}M\\ (eV) \end{tabular} & \begin{tabular}[c]{@{}c@{}}K\\ (eV) \end{tabular} \\ \hline
0.10                                                       & -0.12 &  -0.329 & -0.005  &  -0.003 \\ \hline
0.42                                                       &  -0.046                    &  -1.265 & -0.020  & -0.013  \\ \hline
1.68                                                       &    -0.178                  &  -4.323 &  -0.080 & -0.033  \\ \hline
3.77                                                       &   -0.392                   &  -4.065 & -0.179  &  -0.020 \\ \hline
6.72                                                       &     -0.676                 &  -3.120 &  -0.317 & -0.002  \\ \hline
10.51                                                      &      -1.011                &  -3.858 & -0.494  & -0.006   \\ \hline
\end{tabular}
\caption{Net band energies renormalization calculated upon excitation with a resonant pump pulse at different points ($\Gamma$, A, M, K) of the BZ and for different absorbed pump fluences.}
\end{table}

Remarkably, we find that the pump pulse at 1.55 eV cannot cause substantial renormalization of the electronic structure at the A, K and M points of the BZ. The only exception is represented by the $\Gamma$ point, in which the single-particle energies significantly depend on the absorbed pump fluence. In this regard, the renormalization of the Mg $\zeta$ band has an impact on the interband transition at $\Gamma$ that determines the $c$-axis plasmon energy, leading to its \textit{redshift}. The time-evolution of the B $\pi$ - Mg $\zeta$ interband transition energy at the $\Gamma$ point of the BZ is shown in Fig. \ref{fig:FigS3}(a) for different field strengths. We observe that the single-particle energies undergo a shrinkage of 40 meV upon increasing the absorbed fluence from 0.10 mJ/cm$^2$ to 1.68 mJ/cm$^2$. This effect is consistently observed at early time-delays (\textit{i.e.} the maximum of the response at 170 fs) in our $\Delta$R/R as a function of fluence (Fig. \ref{fig:FigS3}(b)). The plasmon blueshift decreases by 20 meV when the absorbed pump fluence changes from 0.24 mJ/cm$^2$ to 2.40 mJ/cm$^2$, which has a striking quantitative agreement with the shift predicted by our \textit{ab initio} calculations. The shift observed in the experimental spectra has to be interpreted as a net variation of the plasmon energy that results from the combination of two effects: The blueshift induced by the increased carrier density $\widetilde{n}$ and the redshift caused by the many-body renormalization of the electronic structure. In conclusion, the present calculations rule out the possibility of a many-body electronic effect behind the plasmon blueshift in the rise of the response.\\

\section{S6. Temperature-dependent change of the plasma frequencies}

We now address the anomalous blueshift displayed by the a- and c-axis plasma frequencies. The phenomenological argument raised in the main text for describing the blueshift in the dressed c-axis plasma frequency involved an increase in the effective carrier density $\widetilde{n}$ during the nonequilibrium evolution of the system. Providing an explanation to this process is not obvious, since the total carrier density in the system is constant and can just be redistributed energetically by the action of the pump pulse. As we discussed in the main text, here we demonstrate that the time evolution of the plasma frequency $\omega_{p,a}$ can be attributed to the different role played by the photoinduced heating of the electronic and phononic bath after the excitation. To get a deeper insight on this mechanism, we consider the processes occurring in a multiband system with strongly particle-hole asymmetric bands. Indeed, similarly to the case of pnictides \cite{ortenzi_prl09, benfatto_prb11}, these conditions can make the effect of the electron-phonon coupling visible in an interaction-dependent change of size of the Fermi surface with respect to Local Density Approximation (LDA) calculations. More specifically, for a purely parabolic two-dimensional band, one can relate \cite{ortenzi_prl09, benfatto_prb11} the density of carriers $\tilde{n}$ to the top/bottom $E_M$ of the band as 
\begin{equation}
\label{eqn}
\tilde n=\frac{k_F^2}{2\pi}=\frac{2m(E_M+\chi)}{2\pi}
\end{equation}
where $\chi$ denotes the real part of the electron-phonon self-energy at zero energy. While the Eliashberg calculation of the self-energy gives $\chi=0$ for half-filled bands, at low carrier density $\chi$ becomes in general different from zero, and can lead to an observable shrinking or expansion of the Fermi surface area as a function of the electron-phonon interaction. By computing $\chi$ for an Einstein phonon and making the analytical continuation with the Marsiglio-Schlossmann-Carbotte method \cite{ref:marsiglio_iterative}, one can show that the electronic $\chi_{\rm el}$ and bosonic $\chi_{\rm bos}$ contributions to $\chi$ can be written as
\begin{eqnarray}
\chi_{\rm el}(T=0)
&=&-\lambda
\frac{\omega_0}{2}
\ln\left|
\frac{E_B-\omega_0}
{E_B+\omega_0}
\right|,
\\
\chi_{\rm ph}(T=0)
&=&
-
\lambda
\frac{\omega_0}{2}
\ln\left|
\frac{E_T+\omega_0}
{E_B-\omega_0}
\right|,
\end{eqnarray}
so that
\begin{eqnarray}
\chi_{\rm tot}(T=0)
&=&
\chi_{\rm el}(T=0)+\chi_{\rm ph}(T=0)
\nonumber\\
&=&
\label{chitot}
-\lambda
\frac{\omega_0}{2}
\ln\left|
\frac{E_T+\omega_0}
{E_B+\omega_0}
\right| .
\end{eqnarray}
in agreement with previous estimates \cite{benfatto_prb11}. Here $\omega_0$ is the phonon energy, $\lambda$ is the dimensionless electron-phonon coupling and $E_{T/B}$ represent the band top/bottom, respectively. In a multiband system the previous equation (\ref{chitot}) must be generalized to account for both the interband and intraband contributions, mediated in general by phonons or other collective modes, so that $\chi_{\alpha}=\sum_\beta \lambda_{\alpha\beta} R_\beta$ where $R_\beta=\frac{\omega_0}{2}\ln\left|
\frac{E^\beta_T+\omega_0}{E^\beta_B+\omega_0}\right|$. Since $R_\beta>0$ for a band having electron character (so that $E^\beta_T\gg E^\beta_B$) while $R_\beta<0$ for a band having hole character the final sign of $\chi_\alpha$ depends on the intraband \textit{vs} interband nature of the dominant interaction. For example, in the case of pnictides, the dominant channel is provided by spin-fluctuations interband interactions between hole and electron bands, leading to a shrinking of all the Fermi pockets with respect to LDA \cite{ortenzi_prl09, benfatto_prb11}. In the case of MgB$_2$, we expect that the largest channel of interaction is the intraband coupling between the $\sigma$ electrons and the $\mathrm{E_{2g}}$ phonon. In this situation, we expect that $\chi_\sigma(T=0)>0$, but since the unperturbed Fermi area of the $\sigma$ pocket is much larger than in pnictides this effect is quantitatively much smaller at equilibrium, explaining why in MgB$_2$ angle-resolved photoemission spectroscopy measurements are overall in good agreement with LDA predictions \cite{uchiyama2002electronic}. Nonetheless, as we argued in the main text, it can explain the small variations of the plasma frequency observed by heating the system with an intense laser pulse. 

To address the thermal effects in the situation where the electronic temperature $\mathrm{T_e}$ does not necessarely coincide with the bosonic one $\mathrm{T_b}$ we identify in the expression for $\chi(T)$ two separate terms, one depending on the Fermi function $\mathrm{f(\omega_0,T_e)}$, controlled by the $\mathrm{T_e}$, and one depending on the Bose function $\mathrm{b(\omega_0,T_b)}$, controlled by $\mathrm{T_b}$. This distinction can be achieved by performing the analytical continuation of the Matsubara self-energy $\Sigma(i\omega_n)$ without leaving the ambiguity in the use of the Fermi and Bose factors. One can show that the final result 
at the lowest order of the perturbation theory reads
\begin{eqnarray}
\label{chie}
\Delta\chi_{\rm el}(T) &=& -\lambda \frac{T}{2}\sum_m \frac{\omega_0^2}{\omega_m^2+\omega_0^2}
\ln \left| \frac{E_T^2+\omega_m^2}{E_B^2+\omega_m^2}
\right|
\nonumber\\
&& - \lambda
\frac{\omega_0}{2}
f(\omega_0)
\ln\left|
\frac{E_T^2-\omega_0^2}
{E_B^2-\omega_0^2}
\right|
\nonumber\\
&& + \lambda
\frac{\omega_0}{2}
\ln\left|
\frac{E_T+\omega_0}
{E_B-\omega_0}
\right|
,
\\
\label{chib}
\Delta\chi_{\rm ph}(T)
&=&
-
\lambda
\frac{\omega_0}{2}
b(\omega_0)
\ln\left|
\frac{E_T^2-\omega_0^2}
{E_B^2-\omega_0^2}
\right|
.
\end{eqnarray}

\begin{figure}[t]
\includegraphics[width=0.8\columnwidth]{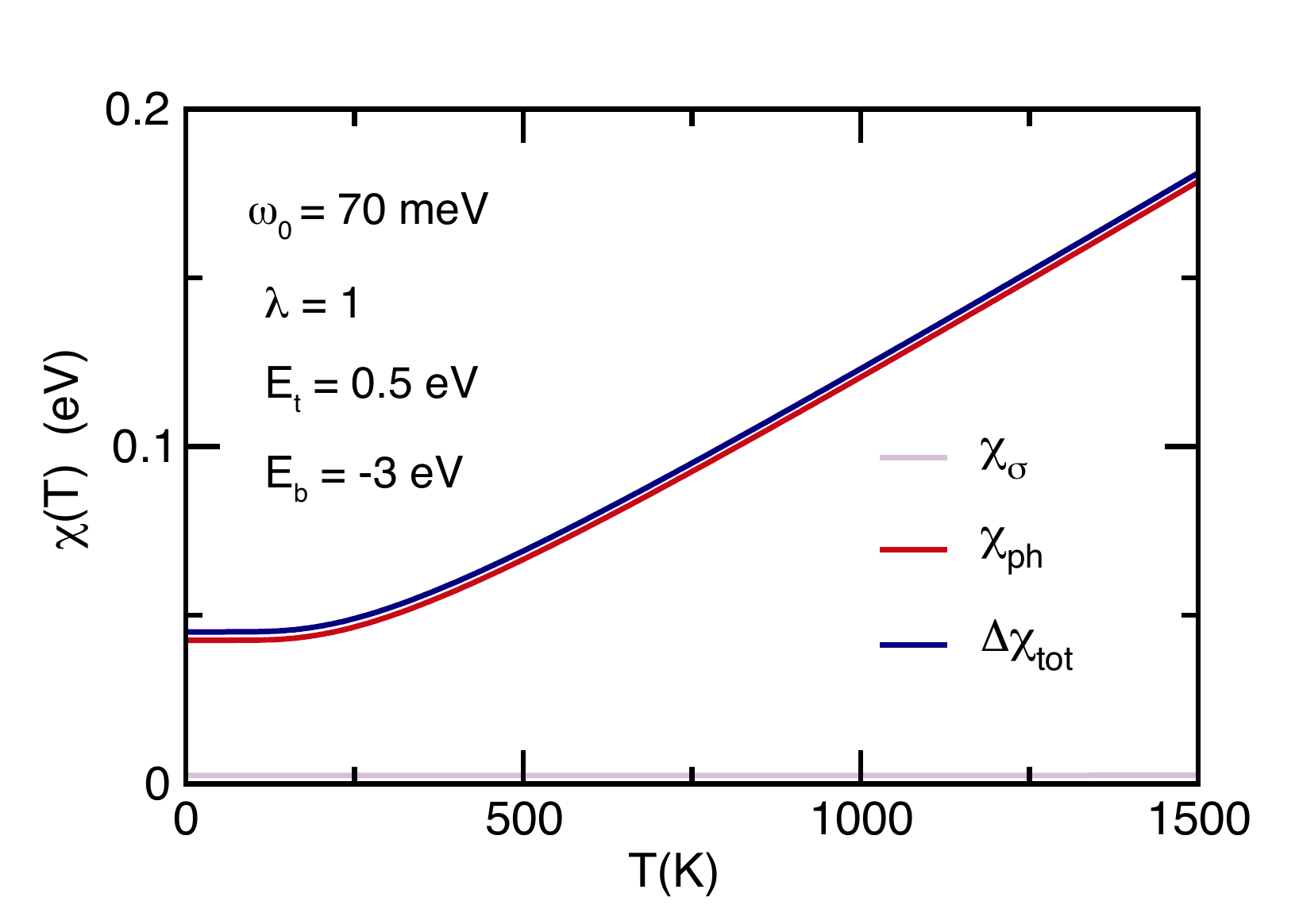}
\caption{Temperature dependence for the bosonic and electronic contributions to the shift of the $\sigma$ bands in temperature}
\label{fig-chi}
\end{figure}

The two separate contributions (\ref{chie})-(\ref{chib}) are computed for realistic parameter values appropriate for the $\sigma$ band. In particular, the phonon energy $\omega_0$ is set at 70 meV to account for the strong peak found at this energy in the Eliashberg function $\alpha^2$ F($\omega$) \cite{ref:bohnen}. This is mainly governed by the contribution of the strongly softened in-plane vibrations of the boron atoms. The result is shown in Fig.\ \ref{fig-chi}, where one clearly sees the predominance of the thermal effects due to heating of the phonon bath. This has a direct impact on the evolution of the plasma frequencies of the $\sigma$ and $\pi$ bands after the pump. Indeed, the initial heating of the electronic bath due to the pump is also effective on the $\mathrm{E_{2g}}$ phonon, which is strongly coupled to the $\sigma$ electrons. This in turn implies an {\em increase} of the Fermi area of the $\sigma$ pocket that is coupled to the $\mathrm{E_{2g}}$ hot phonon, while the $\pi$ band is unaffected, since it is weakly coupled to the remaining cold phononic bath. Notice that the initial increase of the plasma frequency in the $\sigma$ band is in striking contrast with the usual findings in single-band systems, where heating the electronic bath results in a suppression of the plasma edge \cite{toschi_prl05,nicoletti_prl10,kuzmenko_prb07}.

After the first 170 fs, the system equilibrates to a state with hot $\sigma/\pi$ carriers and $\mathrm{E_{2g}}$ phonon, and then starts to equilibrate with the remaining phononic bath with the less efficient electron-phonon scattering channels and the phonon-phonon anharmonic processes. This relaxation process activates the interband $\sigma$-$\pi$ scattering mechanisms which leads to an increase of the $\pi$ Fermi surface, needed to preserve the total number of carriers in the system. After about 2 ps, the system is then overall at equilibrium with a larger effective number of carriers in all bands, as reported by the optical measurements. The above scheme can be used to indirectly estimate the heating of the hot phononic bath via its effect on the bare plasma frequency of the $\sigma$ band, given by 
\begin{equation}
\label{eqp}
\omega_{p,\sigma}=\sqrt{\frac{4\pi e^2 \tilde n_{\sigma}}{m}}
\end{equation}
From Eq.\ \ref{eqp} we can estimate the change of $\tilde n_\sigma$ due to the measured change in $\omega_{p,\sigma}$. Indeed, we have that
\begin{equation}
\label{delta}
\delta_\sigma\equiv\frac{\Delta \omega_{p,\sigma}}{\omega_{p,\sigma}}=\frac{1}{2}\frac{\Delta \tilde n_\sigma}{\tilde n_\sigma}\simeq 1.23\times 10^{-2}
\end{equation}
Since $\Delta\tilde n_\sigma$ depends on the $\Delta\chi_\sigma$ computed above, we then conclude that the hot phonons reach a temperature around 450 K.

\section{S7. Lattice expansion effect on the plasmon energy}

After elucidating the origin of the anomalous blueshift of the a- and c-axis plasma frequencies, here we concentrate on the long timescale dynamics of the system, subsequent to the extinction of hot phonon effects. Our aim is to evaluate how the c-axis plasmon energy is renormalized upon lattice expansion. As discussed in the main text, lattice expansion is expected to result from hot phonon energy dissipation via anharmonic coupling with low-energy acoustic modes. To this end, we perform \textit{ab initio} calculations of the dynamical electronic properties of MgB$_2$ for different lattice parameters in the case of uniaxial (c-axis) and of three-dimensional lattice expansion. Specifically, we simulate the effect of the lattice expansion on the c-axis plasmon of MgB$_2$ by calculating the loss function $L({\bf q},\omega)$ in the long-wavelength limit (\textbf{q} = 0) and at different lattice parameters.

\begin{figure}[b]
\begin{center}
\includegraphics[width=\columnwidth]{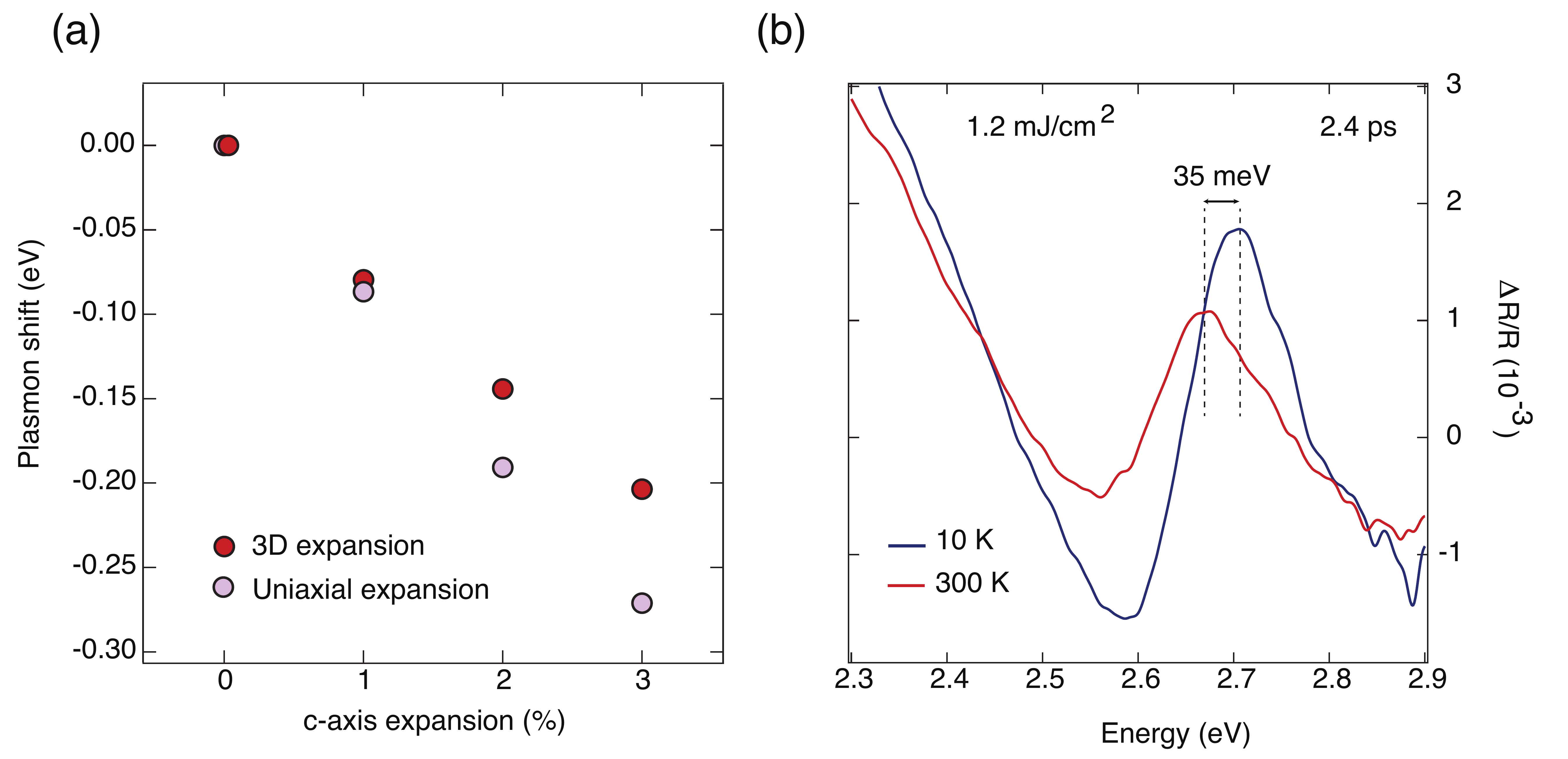}
\caption{(a) Calculated plasmon shift as a function of the crystal expansion along the c-axis, which mimics the effect of thermal heating on the plasmon energy. The values are reported in two different cases: Three-dimensional expansion (red dots) and uniaxial expansion (violet dots). (b) Transient spectra of $\Delta$R/R in the vicinity of the $c$-axis plasmon peak at 10 and 300 K, for an absorbed fluence of 1.2 mJ/cm$^2$ and at a delay time of 2.4 ps.}
\label{fig:Anharmonic}
\end{center}
\end{figure}

Two models for the MgB$_2$ lattice parameter expansion were employed. In the first case, we obtain the lattice parameters by considering a three-dimensional lattice expansion. Here, at a fixed volume value, the c/a ratio is optimized using the Vienna Ab Initio Simulation Package (VASP) \cite{VASP1,VASP2} with Generalized Gradient Approximation (GGA) \cite{ref:perdew} for the exchange-correlation potential. The interaction between the ion cores and valence electrons was described by the projector augmented-wave method \cite{PAW}. In the second case, we perform calculations for the uniaxial $c$ expansion, \textit{i.e.} varying the lattice constant only in this direction and maintaining the in-plane $a$ constant.

The collective electronic excitations of a bulk solid can be traced \cite{pino66,givi08} to the peaks in the loss function $L({\bf q},\omega)$, defined as the imaginary part of the inverse dielectric function
\begin{equation}
L({\bf q},\omega)={\rm Im}[\epsilon^{-1}({\bf q},\omega)],
\end{equation}
where ${\bf q}$ and $\omega$ are momentum and energy, respectively, transferred to the system. The inverse dielectric function $\epsilon^{-1}$ is related to the density-response function of interacting electrons $\chi$  through the integral equation $\epsilon^{-1}=1+\upsilon\chi$, where $\upsilon$ is the Coulomb potential. In the framework of the Time-Dependent Density-Functional Theory \cite{rugrprl84,pegoprl96} $\chi$ obeys the integral equation  $\chi=\chi^o +\chi^o(\upsilon+{\rm K}_{\rm xc})\chi$, where $\chi^o$ is the response function for a non-interacting electron system and K$_{\rm xc}$ accounts for dynamical exchange-correlation effects. The imaginary part of $\chi^o$, which takes a matrix form for a three-dimensional solid, is calculated according to
\begin{widetext}
\begin{equation}\label{spectral_function}
{\rm Im}[\chi^o_{{\bf G}{\bf G}'}({\bf q},\omega)] = \frac{2}{\Omega}
\sum^{\rm BZ}_{\bf k} \sum_{nn'}
(f_{n{\bf k}}-f_{n'{\bf k}+{\bf q}}) \langle\psi_{n{\bf k}}|e^{-{\rm i}({\bf q}+{\bf G})\cdot{\bf
r}}|\psi_{n'{\bf k}+{\bf q}}\rangle \langle\psi_{n'{\bf k}+{\bf
q}}|e^{{\rm i}({\bf q}+{\bf G}')\cdot{\bf r}}|\psi_{n{\bf k}}\rangle
\delta(\varepsilon_{n{\bf k}}-\varepsilon_{n'{\bf k}+{\bf
q}}+\omega),
\end{equation}
\end{widetext}
where the factor 2 accounts for spin, $\Omega$ is the normalization volume, $n$ and $n'$ are the energy band indices, vector ${\bf k}$ is in the first BZ, $f_{n{\bf k}}$ is the Fermi distribution function, $\varepsilon_{n{\bf k}}$  and $\psi_{n{\bf k}}$ are Bloch eigenvalues and eigenfunctions, respectively, of the Kohn-Sham Hamiltonian. In the numerical calculations, the $\delta$-function in Eq. (\ref{spectral_function}) is replaced by a Gaussian with a broadening parameter of 25 meV. Subsequently, the real part of $\chi^o$ is obtained from Im$[\chi^o]$ (evaluated on a discrete mesh of energies ranging from 0 to 25 eV with the step of 2.5 meV) using the Kramers-Kronig relation. The single-particle energies and wave functions are obtained from the self-consistent solution of the Kohn-Sham equations using the local exchange-correlation potential of Refs. \onlinecite{cealprl80,pezuprb81}. The electron-ion interaction is described by a nonlocal norm-conserving ionic pseudopotential \cite{trmaprb91}. In the Fourier expansion of $\chi^o$, $\chi$ and $\epsilon$, matrices up to 51 reciprocal lattice vectors ${\bf G}$ are included, in such a way to account for the local-field effects \cite{adpr62,wipr63}. In equation \ref{spectral_function}, we use a (120 $\times$ 120 $\times$ 60) mesh for the ${\bf k}$ summation over the first BZ. The sum over $n$ and $n'$ includes 30 valence bands. For the description of K$_{\rm xc}$ a Random-Phase Approximation (\textit{i.e.} K$_{\rm xc}$ = 0) was employed.

In both analyzed cases, we find that the c-axis plasmon peak energy redshifts by 37 meV/P, where P is rate of the volume expansion in percent. The results of the calculations are shown in Fig. \ref{fig:Anharmonic}(a). The origin of such behavior resides again in the fact that the plasmon energy is mainly determined by the transitions from the B $\pi$ band to the Mg $\zeta$ band at the $\Gamma$ point of the BZ. As the distance between atomic planes increases, the energy separation between these two bands reduces, which results in the plasmon \textit{redshift}. This scenario is confirmed by our $\Delta$R/R data at 10 K and 300 K (Fig. \ref{fig:Anharmonic}(b)) at a time delay of 2.4 ps, \textit{i.e.} well after the extinction of hot phonon effects. At this time delay, we observe a redshift of $\sim$ 35 meV in the $c$-axis plasmon peak when the temperature is increased from 10 K to 300 K, due to the increased lattice constants in the crystal. This result rationalizes the last step of the dynamics probed within our temporal window and offers a complete explanation of the ultrafast optical response of MgB$_2$.
\clearpage
\newpage

\bibliography{Papers}

\providecommand{\noopsort}[1]{}\providecommand{\singleletter}[1]{#1}%
\begin{thebibliography}{67}
\expandafter\ifx\csname natexlab\endcsname\relax\def\natexlab#1{#1}\fi
\expandafter\ifx\csname bibnamefont\endcsname\relax
  \def\bibnamefont#1{#1}\fi
\expandafter\ifx\csname bibfnamefont\endcsname\relax
  \def\bibfnamefont#1{#1}\fi
\expandafter\ifx\csname citenamefont\endcsname\relax
  \def\citenamefont#1{#1}\fi
\expandafter\ifx\csname url\endcsname\relax
  \def\url#1{\texttt{#1}}\fi
\expandafter\ifx\csname urlprefix\endcsname\relax\def\urlprefix{URL }\fi
\providecommand{\bibinfo}[2]{#2}
\providecommand{\eprint}[2][]{\url{#2}}

\bibitem[{\citenamefont{Dresselhaus and
  Dresselhaus}(1981)}]{dresselhaus1981intercalation}
\bibinfo{author}{\bibfnamefont{M.}~\bibnamefont{Dresselhaus}} \bibnamefont{and}
  \bibinfo{author}{\bibfnamefont{G.}~\bibnamefont{Dresselhaus}},
  \bibinfo{journal}{Advances in Physics} \textbf{\bibinfo{volume}{30}},
  \bibinfo{pages}{139} (\bibinfo{year}{1981}).

\bibitem[{\citenamefont{Hebard et~al.}(1991)\citenamefont{Hebard, Rosseinky,
  Haddon, Murphy, Glarum, Palstra, Ramirez, and Karton}}]{hebard1991potassium}
\bibinfo{author}{\bibfnamefont{A.}~\bibnamefont{Hebard}},
  \bibinfo{author}{\bibfnamefont{M.}~\bibnamefont{Rosseinky}},
  \bibinfo{author}{\bibfnamefont{R.}~\bibnamefont{Haddon}},
  \bibinfo{author}{\bibfnamefont{D.}~\bibnamefont{Murphy}},
  \bibinfo{author}{\bibfnamefont{S.}~\bibnamefont{Glarum}},
  \bibinfo{author}{\bibfnamefont{T.}~\bibnamefont{Palstra}},
  \bibinfo{author}{\bibfnamefont{A.}~\bibnamefont{Ramirez}}, \bibnamefont{and}
  \bibinfo{author}{\bibfnamefont{A.}~\bibnamefont{Karton}},
  \bibinfo{journal}{Nature} \textbf{\bibinfo{volume}{350}},
  \bibinfo{pages}{600} (\bibinfo{year}{1991}).

\bibitem[{\citenamefont{Bednorz and M{\"u}ller}(1986)}]{bednorz1986possible}
\bibinfo{author}{\bibfnamefont{J.~G.} \bibnamefont{Bednorz}} \bibnamefont{and}
  \bibinfo{author}{\bibfnamefont{K.~A.} \bibnamefont{M{\"u}ller}}, in
  \emph{\bibinfo{booktitle}{Ten Years of Superconductivity: 1980--1990}}
  (\bibinfo{publisher}{Springer}, \bibinfo{year}{1986}), pp.
  \bibinfo{pages}{267--271}.

\bibitem[{\citenamefont{Yu et~al.}(1991)\citenamefont{Yu, Lee, Heeger, Herron,
  and McCarron}}]{yu1991transient}
\bibinfo{author}{\bibfnamefont{G.}~\bibnamefont{Yu}},
  \bibinfo{author}{\bibfnamefont{C.}~\bibnamefont{Lee}},
  \bibinfo{author}{\bibfnamefont{A.}~\bibnamefont{Heeger}},
  \bibinfo{author}{\bibfnamefont{N.}~\bibnamefont{Herron}}, \bibnamefont{and}
  \bibinfo{author}{\bibfnamefont{E.}~\bibnamefont{McCarron}},
  \bibinfo{journal}{Physical review letters} \textbf{\bibinfo{volume}{67}},
  \bibinfo{pages}{2581} (\bibinfo{year}{1991}).

\bibitem[{\citenamefont{Mankowsky et~al.}(2016)\citenamefont{Mankowsky,
  F{\"o}rst, and Cavalleri}}]{mankowsky2016non}
\bibinfo{author}{\bibfnamefont{R.}~\bibnamefont{Mankowsky}},
  \bibinfo{author}{\bibfnamefont{M.}~\bibnamefont{F{\"o}rst}},
  \bibnamefont{and}
  \bibinfo{author}{\bibfnamefont{A.}~\bibnamefont{Cavalleri}},
  \bibinfo{journal}{Reports on Progress in Physics}
  \textbf{\bibinfo{volume}{79}}, \bibinfo{pages}{064503}
  (\bibinfo{year}{2016}).

\bibitem[{\citenamefont{Kampfrath et~al.}(2005)\citenamefont{Kampfrath,
  Perfetti, Schapper, Frischkorn, and Wolf}}]{ref:kampfrath}
\bibinfo{author}{\bibfnamefont{T.}~\bibnamefont{Kampfrath}},
  \bibinfo{author}{\bibfnamefont{L.}~\bibnamefont{Perfetti}},
  \bibinfo{author}{\bibfnamefont{F.}~\bibnamefont{Schapper}},
  \bibinfo{author}{\bibfnamefont{C.}~\bibnamefont{Frischkorn}},
  \bibnamefont{and} \bibinfo{author}{\bibfnamefont{M.}~\bibnamefont{Wolf}},
  \bibinfo{journal}{Phys. Rev. Lett.} \textbf{\bibinfo{volume}{95}},
  \bibinfo{pages}{187403} (\bibinfo{year}{2005}).

\bibitem[{\citenamefont{Yan et~al.}(2009)\citenamefont{Yan, Song, Mak,
  Chatzakis, Maultzsch, and Heinz}}]{heinz_prb09}
\bibinfo{author}{\bibfnamefont{H.}~\bibnamefont{Yan}},
  \bibinfo{author}{\bibfnamefont{D.}~\bibnamefont{Song}},
  \bibinfo{author}{\bibfnamefont{K.~F.} \bibnamefont{Mak}},
  \bibinfo{author}{\bibfnamefont{I.}~\bibnamefont{Chatzakis}},
  \bibinfo{author}{\bibfnamefont{J.}~\bibnamefont{Maultzsch}},
  \bibnamefont{and} \bibinfo{author}{\bibfnamefont{T.~F.} \bibnamefont{Heinz}},
  \bibinfo{journal}{Phys. Rev. B} \textbf{\bibinfo{volume}{80}},
  \bibinfo{pages}{121403} (\bibinfo{year}{2009}).

\bibitem[{\citenamefont{Breusing et~al.}(2009)\citenamefont{Breusing, Ropers,
  and Elsaesser}}]{ref:breusing}
\bibinfo{author}{\bibfnamefont{M.}~\bibnamefont{Breusing}},
  \bibinfo{author}{\bibfnamefont{C.}~\bibnamefont{Ropers}}, \bibnamefont{and}
  \bibinfo{author}{\bibfnamefont{T.}~\bibnamefont{Elsaesser}},
  \bibinfo{journal}{Phys. Rev. Lett.} \textbf{\bibinfo{volume}{102}},
  \bibinfo{pages}{086809} (\bibinfo{year}{2009}).

\bibitem[{\citenamefont{Lui et~al.}(2010)\citenamefont{Lui, Mak, Shan, and
  Heinz}}]{heinz_prl10}
\bibinfo{author}{\bibfnamefont{C.~H.} \bibnamefont{Lui}},
  \bibinfo{author}{\bibfnamefont{K.~F.} \bibnamefont{Mak}},
  \bibinfo{author}{\bibfnamefont{J.}~\bibnamefont{Shan}}, \bibnamefont{and}
  \bibinfo{author}{\bibfnamefont{T.~F.} \bibnamefont{Heinz}},
  \bibinfo{journal}{Phys. Rev. Lett.} \textbf{\bibinfo{volume}{105}},
  \bibinfo{pages}{127404} (\bibinfo{year}{2010}).

\bibitem[{\citenamefont{Breusing et~al.}(2011)\citenamefont{Breusing, Kuehn,
  Winzer, Mali{\'c}, Milde, Severin, Rabe, Ropers, Knorr, and
  Elsaesser}}]{ref:breusing2011}
\bibinfo{author}{\bibfnamefont{M.}~\bibnamefont{Breusing}},
  \bibinfo{author}{\bibfnamefont{S.}~\bibnamefont{Kuehn}},
  \bibinfo{author}{\bibfnamefont{T.}~\bibnamefont{Winzer}},
  \bibinfo{author}{\bibfnamefont{E.}~\bibnamefont{Mali{\'c}}},
  \bibinfo{author}{\bibfnamefont{F.}~\bibnamefont{Milde}},
  \bibinfo{author}{\bibfnamefont{N.}~\bibnamefont{Severin}},
  \bibinfo{author}{\bibfnamefont{J.~P.} \bibnamefont{Rabe}},
  \bibinfo{author}{\bibfnamefont{C.}~\bibnamefont{Ropers}},
  \bibinfo{author}{\bibfnamefont{A.}~\bibnamefont{Knorr}}, \bibnamefont{and}
  \bibinfo{author}{\bibfnamefont{T.}~\bibnamefont{Elsaesser}},
  \bibinfo{journal}{Phys. Rev. B} \textbf{\bibinfo{volume}{83}},
  \bibinfo{pages}{153410} (\bibinfo{year}{2011}).

\bibitem[{\citenamefont{Dal~Conte et~al.}(2015)\citenamefont{Dal~Conte, Vidmar,
  Gole{\v{z}}, Mierzejewski, Soavi, Peli, Banfi, Ferrini, Comin, Ludbrook
  et~al.}}]{ref:dalconte}
\bibinfo{author}{\bibfnamefont{S.}~\bibnamefont{Dal~Conte}},
  \bibinfo{author}{\bibfnamefont{L.}~\bibnamefont{Vidmar}},
  \bibinfo{author}{\bibfnamefont{D.}~\bibnamefont{Gole{\v{z}}}},
  \bibinfo{author}{\bibfnamefont{M.}~\bibnamefont{Mierzejewski}},
  \bibinfo{author}{\bibfnamefont{G.}~\bibnamefont{Soavi}},
  \bibinfo{author}{\bibfnamefont{S.}~\bibnamefont{Peli}},
  \bibinfo{author}{\bibfnamefont{F.}~\bibnamefont{Banfi}},
  \bibinfo{author}{\bibfnamefont{G.}~\bibnamefont{Ferrini}},
  \bibinfo{author}{\bibfnamefont{R.}~\bibnamefont{Comin}},
  \bibinfo{author}{\bibfnamefont{B.~M.} \bibnamefont{Ludbrook}},
  \bibnamefont{et~al.}, \bibinfo{journal}{Nat. Phys.}
  \textbf{\bibinfo{volume}{11}}, \bibinfo{pages}{421} (\bibinfo{year}{2015}).

\bibitem[{\citenamefont{Nagamatsu et~al.}(2001)\citenamefont{Nagamatsu,
  Nakagawa, Muranaka, Zenitani, and Akimitsu}}]{ref:nagamatsu}
\bibinfo{author}{\bibfnamefont{J.}~\bibnamefont{Nagamatsu}},
  \bibinfo{author}{\bibfnamefont{N.}~\bibnamefont{Nakagawa}},
  \bibinfo{author}{\bibfnamefont{T.}~\bibnamefont{Muranaka}},
  \bibinfo{author}{\bibfnamefont{Y.}~\bibnamefont{Zenitani}}, \bibnamefont{and}
  \bibinfo{author}{\bibfnamefont{J.}~\bibnamefont{Akimitsu}},
  \bibinfo{journal}{Nature (London)} \textbf{\bibinfo{volume}{410}},
  \bibinfo{pages}{63} (\bibinfo{year}{2001}).

\bibitem[{\citenamefont{Kong et~al.}(2001)\citenamefont{Kong, Dolgov, Jepsen,
  and Andersen}}]{ref:kong}
\bibinfo{author}{\bibfnamefont{Y.}~\bibnamefont{Kong}},
  \bibinfo{author}{\bibfnamefont{O.}~\bibnamefont{Dolgov}},
  \bibinfo{author}{\bibfnamefont{O.}~\bibnamefont{Jepsen}}, \bibnamefont{and}
  \bibinfo{author}{\bibfnamefont{O.}~\bibnamefont{Andersen}},
  \bibinfo{journal}{Phys. Rev. B} \textbf{\bibinfo{volume}{64}},
  \bibinfo{pages}{020501} (\bibinfo{year}{2001}).

\bibitem[{\citenamefont{Yelland et~al.}(2002)\citenamefont{Yelland, Cooper,
  Carrington, Hussey, Meeson, Lee, Yamamoto, and Tajima}}]{yelland}
\bibinfo{author}{\bibfnamefont{E.~A.} \bibnamefont{Yelland}},
  \bibinfo{author}{\bibfnamefont{J.~R.} \bibnamefont{Cooper}},
  \bibinfo{author}{\bibfnamefont{A.}~\bibnamefont{Carrington}},
  \bibinfo{author}{\bibfnamefont{N.~E.} \bibnamefont{Hussey}},
  \bibinfo{author}{\bibfnamefont{P.~J.} \bibnamefont{Meeson}},
  \bibinfo{author}{\bibfnamefont{S.}~\bibnamefont{Lee}},
  \bibinfo{author}{\bibfnamefont{A.}~\bibnamefont{Yamamoto}}, \bibnamefont{and}
  \bibinfo{author}{\bibfnamefont{S.}~\bibnamefont{Tajima}},
  \bibinfo{journal}{Phys. Rev. Lett.} \textbf{\bibinfo{volume}{88}},
  \bibinfo{pages}{217002} (\bibinfo{year}{2002}).

\bibitem[{\citenamefont{Choi et~al.}(2002)\citenamefont{Choi, Roundy, Sun,
  Cohen, and Louie}}]{ref:choi}
\bibinfo{author}{\bibfnamefont{H.~J.} \bibnamefont{Choi}},
  \bibinfo{author}{\bibfnamefont{D.}~\bibnamefont{Roundy}},
  \bibinfo{author}{\bibfnamefont{H.}~\bibnamefont{Sun}},
  \bibinfo{author}{\bibfnamefont{M.~L.} \bibnamefont{Cohen}}, \bibnamefont{and}
  \bibinfo{author}{\bibfnamefont{S.~G.} \bibnamefont{Louie}},
  \bibinfo{journal}{Nature (London)} \textbf{\bibinfo{volume}{418}},
  \bibinfo{pages}{758} (\bibinfo{year}{2002}).

\bibitem[{\citenamefont{Golubov et~al.}(2002)\citenamefont{Golubov, Kortus,
  Dolgov, Jepsen, Kong, Andersen, Gibson, Ahn, and Kremer}}]{ref:golubov}
\bibinfo{author}{\bibfnamefont{A.}~\bibnamefont{Golubov}},
  \bibinfo{author}{\bibfnamefont{J.}~\bibnamefont{Kortus}},
  \bibinfo{author}{\bibfnamefont{O.}~\bibnamefont{Dolgov}},
  \bibinfo{author}{\bibfnamefont{O.}~\bibnamefont{Jepsen}},
  \bibinfo{author}{\bibfnamefont{Y.}~\bibnamefont{Kong}},
  \bibinfo{author}{\bibfnamefont{O.}~\bibnamefont{Andersen}},
  \bibinfo{author}{\bibfnamefont{B.}~\bibnamefont{Gibson}},
  \bibinfo{author}{\bibfnamefont{K.}~\bibnamefont{Ahn}}, \bibnamefont{and}
  \bibinfo{author}{\bibfnamefont{R.}~\bibnamefont{Kremer}},
  \bibinfo{journal}{J. Phys. Cond. Matt.} \textbf{\bibinfo{volume}{14}},
  \bibinfo{pages}{1353} (\bibinfo{year}{2002}).

\bibitem[{\citenamefont{Souma et~al.}(2003)\citenamefont{Souma, Machida, Sato,
  Takahashi, Matsui, Wang, Ding, Kaminski, Campuzano, Sasaki
  et~al.}}]{ref:souma}
\bibinfo{author}{\bibfnamefont{S.}~\bibnamefont{Souma}},
  \bibinfo{author}{\bibfnamefont{Y.}~\bibnamefont{Machida}},
  \bibinfo{author}{\bibfnamefont{T.}~\bibnamefont{Sato}},
  \bibinfo{author}{\bibfnamefont{T.}~\bibnamefont{Takahashi}},
  \bibinfo{author}{\bibfnamefont{H.}~\bibnamefont{Matsui}},
  \bibinfo{author}{\bibfnamefont{S.-C.} \bibnamefont{Wang}},
  \bibinfo{author}{\bibfnamefont{H.}~\bibnamefont{Ding}},
  \bibinfo{author}{\bibfnamefont{A.}~\bibnamefont{Kaminski}},
  \bibinfo{author}{\bibfnamefont{J.}~\bibnamefont{Campuzano}},
  \bibinfo{author}{\bibfnamefont{S.}~\bibnamefont{Sasaki}},
  \bibnamefont{et~al.}, \bibinfo{journal}{Nature (London)}
  \textbf{\bibinfo{volume}{423}}, \bibinfo{pages}{65} (\bibinfo{year}{2003}).

\bibitem[{\citenamefont{Quilty et~al.}(2002)\citenamefont{Quilty, Lee,
  Yamamoto, and Tajima}}]{ref:quilty}
\bibinfo{author}{\bibfnamefont{J.}~\bibnamefont{Quilty}},
  \bibinfo{author}{\bibfnamefont{S.}~\bibnamefont{Lee}},
  \bibinfo{author}{\bibfnamefont{A.}~\bibnamefont{Yamamoto}}, \bibnamefont{and}
  \bibinfo{author}{\bibfnamefont{S.}~\bibnamefont{Tajima}},
  \bibinfo{journal}{Phys. Rev. Lett.} \textbf{\bibinfo{volume}{88}},
  \bibinfo{pages}{087001} (\bibinfo{year}{2002}).

\bibitem[{\citenamefont{Guritanu et~al.}(2006)\citenamefont{Guritanu, Kuzmenko,
  van~der Marel, Kazakov, Zhigadlo, and Karpinski}}]{ref:guritanu}
\bibinfo{author}{\bibfnamefont{V.}~\bibnamefont{Guritanu}},
  \bibinfo{author}{\bibfnamefont{A.~B.} \bibnamefont{Kuzmenko}},
  \bibinfo{author}{\bibfnamefont{D.}~\bibnamefont{van~der Marel}},
  \bibinfo{author}{\bibfnamefont{S.}~\bibnamefont{Kazakov}},
  \bibinfo{author}{\bibfnamefont{N.}~\bibnamefont{Zhigadlo}}, \bibnamefont{and}
  \bibinfo{author}{\bibfnamefont{J.}~\bibnamefont{Karpinski}},
  \bibinfo{journal}{Phys. Rev. B} \textbf{\bibinfo{volume}{73}},
  \bibinfo{pages}{104509} (\bibinfo{year}{2006}).

\bibitem[{\citenamefont{Kakeshita et~al.}(2006)\citenamefont{Kakeshita, Lee,
  and Tajima}}]{ref:kakeshita}
\bibinfo{author}{\bibfnamefont{T.}~\bibnamefont{Kakeshita}},
  \bibinfo{author}{\bibfnamefont{S.}~\bibnamefont{Lee}}, \bibnamefont{and}
  \bibinfo{author}{\bibfnamefont{S.}~\bibnamefont{Tajima}},
  \bibinfo{journal}{Phys. Rev. Lett.} \textbf{\bibinfo{volume}{97}},
  \bibinfo{pages}{037002} (\bibinfo{year}{2006}).

\bibitem[{\citenamefont{Toschi et~al.}(2005)\citenamefont{Toschi, Capone,
  Ortolani, Calvani, Lupi, and Castellani}}]{toschi_prl05}
\bibinfo{author}{\bibfnamefont{A.}~\bibnamefont{Toschi}},
  \bibinfo{author}{\bibfnamefont{M.}~\bibnamefont{Capone}},
  \bibinfo{author}{\bibfnamefont{M.}~\bibnamefont{Ortolani}},
  \bibinfo{author}{\bibfnamefont{P.}~\bibnamefont{Calvani}},
  \bibinfo{author}{\bibfnamefont{S.}~\bibnamefont{Lupi}}, \bibnamefont{and}
  \bibinfo{author}{\bibfnamefont{C.}~\bibnamefont{Castellani}},
  \bibinfo{journal}{Phys. Rev. Lett.} \textbf{\bibinfo{volume}{95}},
  \bibinfo{pages}{097002} (\bibinfo{year}{2005}).

\bibitem[{\citenamefont{van Heumen et~al.}(2007)\citenamefont{van Heumen,
  Lortz, Kuzmenko, Carbone, van~der Marel, Zhao, Yu, Cho, Barisic, Greven
  et~al.}}]{kuzmenko_prb07}
\bibinfo{author}{\bibfnamefont{E.}~\bibnamefont{van Heumen}},
  \bibinfo{author}{\bibfnamefont{R.}~\bibnamefont{Lortz}},
  \bibinfo{author}{\bibfnamefont{A.~B.} \bibnamefont{Kuzmenko}},
  \bibinfo{author}{\bibfnamefont{F.}~\bibnamefont{Carbone}},
  \bibinfo{author}{\bibfnamefont{D.}~\bibnamefont{van~der Marel}},
  \bibinfo{author}{\bibfnamefont{X.}~\bibnamefont{Zhao}},
  \bibinfo{author}{\bibfnamefont{G.}~\bibnamefont{Yu}},
  \bibinfo{author}{\bibfnamefont{Y.}~\bibnamefont{Cho}},
  \bibinfo{author}{\bibfnamefont{N.}~\bibnamefont{Barisic}},
  \bibinfo{author}{\bibfnamefont{M.}~\bibnamefont{Greven}},
  \bibnamefont{et~al.}, \bibinfo{journal}{Phys. Rev. B}
  \textbf{\bibinfo{volume}{75}}, \bibinfo{pages}{054522}
  (\bibinfo{year}{2007}).

\bibitem[{\citenamefont{Xu et~al.}(2003)\citenamefont{Xu, Khafizov,
  Satrapinsky, K{\'u}{\v{s}}, Plecenik, and Sobolewski}}]{ref:xu}
\bibinfo{author}{\bibfnamefont{Y.}~\bibnamefont{Xu}},
  \bibinfo{author}{\bibfnamefont{M.}~\bibnamefont{Khafizov}},
  \bibinfo{author}{\bibfnamefont{L.}~\bibnamefont{Satrapinsky}},
  \bibinfo{author}{\bibfnamefont{P.}~\bibnamefont{K{\'u}{\v{s}}}},
  \bibinfo{author}{\bibfnamefont{A.}~\bibnamefont{Plecenik}}, \bibnamefont{and}
  \bibinfo{author}{\bibfnamefont{R.}~\bibnamefont{Sobolewski}},
  \bibinfo{journal}{Phys. Rev. Lett.} \textbf{\bibinfo{volume}{91}},
  \bibinfo{pages}{197004} (\bibinfo{year}{2003}).

\bibitem[{\citenamefont{Demsar et~al.}(2003)\citenamefont{Demsar, Averitt,
  Taylor, Kabanov, Kang, Kim, Choi, and Lee}}]{ref:demsar}
\bibinfo{author}{\bibfnamefont{J.}~\bibnamefont{Demsar}},
  \bibinfo{author}{\bibfnamefont{R.}~\bibnamefont{Averitt}},
  \bibinfo{author}{\bibfnamefont{A.}~\bibnamefont{Taylor}},
  \bibinfo{author}{\bibfnamefont{V.}~\bibnamefont{Kabanov}},
  \bibinfo{author}{\bibfnamefont{W.}~\bibnamefont{Kang}},
  \bibinfo{author}{\bibfnamefont{H.}~\bibnamefont{Kim}},
  \bibinfo{author}{\bibfnamefont{E.}~\bibnamefont{Choi}}, \bibnamefont{and}
  \bibinfo{author}{\bibfnamefont{S.}~\bibnamefont{Lee}},
  \bibinfo{journal}{Phys. Rev. Lett.} \textbf{\bibinfo{volume}{91}},
  \bibinfo{pages}{267002} (\bibinfo{year}{2003}).

\bibitem[{\citenamefont{Balassis et~al.}(2008)\citenamefont{Balassis, Chulkov,
  Echenique, and Silkin}}]{ref:balassis}
\bibinfo{author}{\bibfnamefont{A.}~\bibnamefont{Balassis}},
  \bibinfo{author}{\bibfnamefont{E.}~\bibnamefont{Chulkov}},
  \bibinfo{author}{\bibfnamefont{P.}~\bibnamefont{Echenique}},
  \bibnamefont{and} \bibinfo{author}{\bibfnamefont{V.}~\bibnamefont{Silkin}},
  \bibinfo{journal}{Phys. Rev. B} \textbf{\bibinfo{volume}{78}},
  \bibinfo{pages}{224502} (\bibinfo{year}{2008}).

\bibitem[{\citenamefont{Cai et~al.}(2006)\citenamefont{Cai, Chow, Restrepo,
  Takano, Togano, Kito, Ishii, Chen, Liang, Chen et~al.}}]{ref:cai}
\bibinfo{author}{\bibfnamefont{Y.}~\bibnamefont{Cai}},
  \bibinfo{author}{\bibfnamefont{P.}~\bibnamefont{Chow}},
  \bibinfo{author}{\bibfnamefont{O.}~\bibnamefont{Restrepo}},
  \bibinfo{author}{\bibfnamefont{Y.}~\bibnamefont{Takano}},
  \bibinfo{author}{\bibfnamefont{K.}~\bibnamefont{Togano}},
  \bibinfo{author}{\bibfnamefont{H.}~\bibnamefont{Kito}},
  \bibinfo{author}{\bibfnamefont{H.}~\bibnamefont{Ishii}},
  \bibinfo{author}{\bibfnamefont{C.}~\bibnamefont{Chen}},
  \bibinfo{author}{\bibfnamefont{K.}~\bibnamefont{Liang}},
  \bibinfo{author}{\bibfnamefont{C.}~\bibnamefont{Chen}}, \bibnamefont{et~al.},
  \bibinfo{journal}{Phys. Rev. Lett.} \textbf{\bibinfo{volume}{97}},
  \bibinfo{pages}{176402} (\bibinfo{year}{2006}).

\bibitem[{\citenamefont{Ku et~al.}(2002)\citenamefont{Ku, Pickett, Scalettar,
  and Eguiluz}}]{ref:ku}
\bibinfo{author}{\bibfnamefont{W.}~\bibnamefont{Ku}},
  \bibinfo{author}{\bibfnamefont{W.}~\bibnamefont{Pickett}},
  \bibinfo{author}{\bibfnamefont{R.}~\bibnamefont{Scalettar}},
  \bibnamefont{and} \bibinfo{author}{\bibfnamefont{A.}~\bibnamefont{Eguiluz}},
  \bibinfo{journal}{Phys. Rev. Lett.} \textbf{\bibinfo{volume}{88}},
  \bibinfo{pages}{057001} (\bibinfo{year}{2002}).

\bibitem[{\citenamefont{Coldea et~al.}(2008)\citenamefont{Coldea, Fletcher,
  Carrington, Analytis, Bangura, Chu, Erickson, Fisher, Hussey, and
  McDonald}}]{coldea}
\bibinfo{author}{\bibfnamefont{A.~I.} \bibnamefont{Coldea}},
  \bibinfo{author}{\bibfnamefont{J.~D.} \bibnamefont{Fletcher}},
  \bibinfo{author}{\bibfnamefont{A.}~\bibnamefont{Carrington}},
  \bibinfo{author}{\bibfnamefont{J.~G.} \bibnamefont{Analytis}},
  \bibinfo{author}{\bibfnamefont{A.~F.} \bibnamefont{Bangura}},
  \bibinfo{author}{\bibfnamefont{J.-H.} \bibnamefont{Chu}},
  \bibinfo{author}{\bibfnamefont{A.~S.} \bibnamefont{Erickson}},
  \bibinfo{author}{\bibfnamefont{I.~R.} \bibnamefont{Fisher}},
  \bibinfo{author}{\bibfnamefont{N.~E.} \bibnamefont{Hussey}},
  \bibnamefont{and} \bibinfo{author}{\bibfnamefont{R.~D.}
  \bibnamefont{McDonald}}, \bibinfo{journal}{Phys. Rev. Lett.}
  \textbf{\bibinfo{volume}{101}}, \bibinfo{pages}{216402}
  (\bibinfo{year}{2008}).

\bibitem[{\citenamefont{Ding et~al.}(2011)\citenamefont{Ding, Nakayama,
  Richard, Souma, Sato, Takahashi, Neupane, Xu, Pan, Fedorov
  et~al.}}]{ding2011electronic}
\bibinfo{author}{\bibfnamefont{H.}~\bibnamefont{Ding}},
  \bibinfo{author}{\bibfnamefont{K.}~\bibnamefont{Nakayama}},
  \bibinfo{author}{\bibfnamefont{P.}~\bibnamefont{Richard}},
  \bibinfo{author}{\bibfnamefont{S.}~\bibnamefont{Souma}},
  \bibinfo{author}{\bibfnamefont{T.}~\bibnamefont{Sato}},
  \bibinfo{author}{\bibfnamefont{T.}~\bibnamefont{Takahashi}},
  \bibinfo{author}{\bibfnamefont{M.}~\bibnamefont{Neupane}},
  \bibinfo{author}{\bibfnamefont{Y.}~\bibnamefont{Xu}},
  \bibinfo{author}{\bibfnamefont{Z.}~\bibnamefont{Pan}},
  \bibinfo{author}{\bibfnamefont{A.}~\bibnamefont{Fedorov}},
  \bibnamefont{et~al.}, \bibinfo{journal}{Journal of Physics: Condensed Matter}
  \textbf{\bibinfo{volume}{23}}, \bibinfo{pages}{135701}
  (\bibinfo{year}{2011}).

\bibitem[{\citenamefont{Mazin and Antropov}(2003)}]{ref:mazin}
\bibinfo{author}{\bibfnamefont{I.}~\bibnamefont{Mazin}} \bibnamefont{and}
  \bibinfo{author}{\bibfnamefont{V.}~\bibnamefont{Antropov}},
  \bibinfo{journal}{Physica C: Superconductivity}
  \textbf{\bibinfo{volume}{385}}, \bibinfo{pages}{49} (\bibinfo{year}{2003}).

\bibitem[{\citenamefont{Kuzmenko}(2007)}]{ref:kuzmenko}
\bibinfo{author}{\bibfnamefont{A.~B.} \bibnamefont{Kuzmenko}},
  \bibinfo{journal}{Physica C: Superconductivity}
  \textbf{\bibinfo{volume}{456}}, \bibinfo{pages}{63} (\bibinfo{year}{2007}).

\bibitem[{\citenamefont{Baldini et~al.}(2016)\citenamefont{Baldini, Mann,
  Borroni, Arrell, van Mourik, and Carbone}}]{baldini2016versatile}
\bibinfo{author}{\bibfnamefont{E.}~\bibnamefont{Baldini}},
  \bibinfo{author}{\bibfnamefont{A.}~\bibnamefont{Mann}},
  \bibinfo{author}{\bibfnamefont{S.}~\bibnamefont{Borroni}},
  \bibinfo{author}{\bibfnamefont{C.}~\bibnamefont{Arrell}},
  \bibinfo{author}{\bibfnamefont{F.}~\bibnamefont{van Mourik}},
  \bibnamefont{and} \bibinfo{author}{\bibfnamefont{F.}~\bibnamefont{Carbone}},
  \bibinfo{journal}{Structural Dynamics} \textbf{\bibinfo{volume}{3}},
  \bibinfo{pages}{064301} (\bibinfo{year}{2016}).

\bibitem[{\citenamefont{Allen}(1994)}]{ref:allen}
\bibinfo{author}{\bibfnamefont{R.~E.} \bibnamefont{Allen}},
  \bibinfo{journal}{Phys. Rev. B} \textbf{\bibinfo{volume}{50}},
  \bibinfo{pages}{18629} (\bibinfo{year}{1994}).

\bibitem[{\citenamefont{Graves and Allen}(1998)}]{ref:graves}
\bibinfo{author}{\bibfnamefont{J.}~\bibnamefont{Graves}} \bibnamefont{and}
  \bibinfo{author}{\bibfnamefont{R.}~\bibnamefont{Allen}},
  \bibinfo{journal}{Phys. Rev. B} \textbf{\bibinfo{volume}{58}},
  \bibinfo{pages}{13627} (\bibinfo{year}{1998}).

\bibitem[{\citenamefont{Torralva and Allen}(2002)}]{ref:torralva}
\bibinfo{author}{\bibfnamefont{B.}~\bibnamefont{Torralva}} \bibnamefont{and}
  \bibinfo{author}{\bibfnamefont{R.}~\bibnamefont{Allen}}, \bibinfo{journal}{J.
  Mod. Opt.} \textbf{\bibinfo{volume}{49}}, \bibinfo{pages}{593}
  (\bibinfo{year}{2002}).

\bibitem[{\citenamefont{Dou et~al.}(2003)\citenamefont{Dou, Torralva, and
  Allen}}]{ref:dou}
\bibinfo{author}{\bibfnamefont{Y.}~\bibnamefont{Dou}},
  \bibinfo{author}{\bibfnamefont{B.~R.} \bibnamefont{Torralva}},
  \bibnamefont{and} \bibinfo{author}{\bibfnamefont{R.~E.} \bibnamefont{Allen}},
  \bibinfo{journal}{J. Mod. Opt.} \textbf{\bibinfo{volume}{50}},
  \bibinfo{pages}{2615} (\bibinfo{year}{2003}).

\bibitem[{\citenamefont{Allen et~al.}()\citenamefont{Allen, Dumitrica, and
  Torralva}}]{ref:allen2}
\bibinfo{author}{\bibfnamefont{R.~E.} \bibnamefont{Allen}},
  \bibinfo{author}{\bibfnamefont{T.}~\bibnamefont{Dumitrica}},
  \bibnamefont{and} \bibinfo{author}{\bibfnamefont{B.}~\bibnamefont{Torralva}},
  \emph{\bibinfo{title}{{Ultrafast Physical Processes in Semiconductors, edited
  by K. T. Tsen (Academic, New York, 2000)}}}.

\bibitem[{\citenamefont{Crank and Nicolson}(1947)}]{ref:crank}
\bibinfo{author}{\bibfnamefont{J.}~\bibnamefont{Crank}} \bibnamefont{and}
  \bibinfo{author}{\bibfnamefont{P.}~\bibnamefont{Nicolson}}, in
  \emph{\bibinfo{booktitle}{Mathematical Proceedings of the Cambridge
  Philosophical Society}} (\bibinfo{organization}{Cambridge Univ Press},
  \bibinfo{year}{1947}), vol.~\bibinfo{volume}{43}, pp.
  \bibinfo{pages}{50--67}.

\bibitem[{\citenamefont{Sun and Yang}(2011)}]{ref:zgsun}
\bibinfo{author}{\bibfnamefont{Z.}~\bibnamefont{Sun}} \bibnamefont{and}
  \bibinfo{author}{\bibfnamefont{W.}~\bibnamefont{Yang}}, \bibinfo{journal}{J.
  Chem. Phys.} \textbf{\bibinfo{volume}{134}}, \bibinfo{pages}{041101}
  (\bibinfo{year}{2011}).

\bibitem[{\citenamefont{Acocella et~al.}(2012)\citenamefont{Acocella, Jones,
  and Zerbetto}}]{ref:acocella1}
\bibinfo{author}{\bibfnamefont{A.}~\bibnamefont{Acocella}},
  \bibinfo{author}{\bibfnamefont{G.~A.} \bibnamefont{Jones}}, \bibnamefont{and}
  \bibinfo{author}{\bibfnamefont{F.}~\bibnamefont{Zerbetto}},
  \bibinfo{journal}{J. Phys. Chem. Lett.} \textbf{\bibinfo{volume}{3}},
  \bibinfo{pages}{3610} (\bibinfo{year}{2012}).

\bibitem[{\citenamefont{Acocella
  et~al.}(2010{\natexlab{a}})\citenamefont{Acocella, Carbone, and
  Zerbetto}}]{ref:acocella2}
\bibinfo{author}{\bibfnamefont{A.}~\bibnamefont{Acocella}},
  \bibinfo{author}{\bibfnamefont{F.}~\bibnamefont{Carbone}}, \bibnamefont{and}
  \bibinfo{author}{\bibfnamefont{F.}~\bibnamefont{Zerbetto}},
  \bibinfo{journal}{J. Am. Chem. Soc.} \textbf{\bibinfo{volume}{132}},
  \bibinfo{pages}{12166} (\bibinfo{year}{2010}{\natexlab{a}}).

\bibitem[{\citenamefont{Acocella
  et~al.}(2010{\natexlab{b}})\citenamefont{Acocella, Jones, and
  Zerbetto}}]{ref:acocella3}
\bibinfo{author}{\bibfnamefont{A.}~\bibnamefont{Acocella}},
  \bibinfo{author}{\bibfnamefont{G.~A.} \bibnamefont{Jones}}, \bibnamefont{and}
  \bibinfo{author}{\bibfnamefont{F.}~\bibnamefont{Zerbetto}},
  \bibinfo{journal}{J. Phys. Chem. B} \textbf{\bibinfo{volume}{114}},
  \bibinfo{pages}{4101} (\bibinfo{year}{2010}{\natexlab{b}}).

\bibitem[{\citenamefont{Jones et~al.}(2008)\citenamefont{Jones, Acocella, and
  Zerbetto}}]{ref:jones1}
\bibinfo{author}{\bibfnamefont{G.~A.} \bibnamefont{Jones}},
  \bibinfo{author}{\bibfnamefont{A.}~\bibnamefont{Acocella}}, \bibnamefont{and}
  \bibinfo{author}{\bibfnamefont{F.}~\bibnamefont{Zerbetto}},
  \bibinfo{journal}{J. Phys. Chem. A} \textbf{\bibinfo{volume}{112}},
  \bibinfo{pages}{9650} (\bibinfo{year}{2008}).

\bibitem[{\citenamefont{Jones et~al.}(2007)\citenamefont{Jones, Acocella, and
  Zerbetto}}]{ref:jones2}
\bibinfo{author}{\bibfnamefont{G.~A.} \bibnamefont{Jones}},
  \bibinfo{author}{\bibfnamefont{A.}~\bibnamefont{Acocella}}, \bibnamefont{and}
  \bibinfo{author}{\bibfnamefont{F.}~\bibnamefont{Zerbetto}},
  \bibinfo{journal}{Theor. Chem. Acc.} \textbf{\bibinfo{volume}{118}},
  \bibinfo{pages}{99} (\bibinfo{year}{2007}).

\bibitem[{\citenamefont{Acocella et~al.}(2006)\citenamefont{Acocella, Jones,
  and Zerbetto}}]{ref:acocella4}
\bibinfo{author}{\bibfnamefont{A.}~\bibnamefont{Acocella}},
  \bibinfo{author}{\bibfnamefont{G.~A.} \bibnamefont{Jones}}, \bibnamefont{and}
  \bibinfo{author}{\bibfnamefont{F.}~\bibnamefont{Zerbetto}},
  \bibinfo{journal}{J. Phys. Chem. A} \textbf{\bibinfo{volume}{110}},
  \bibinfo{pages}{5164} (\bibinfo{year}{2006}).

\bibitem[{\citenamefont{Acocella et~al.}(2016)\citenamefont{Acocella,
  de~Simone, Evangelista, Coreno, Rudolf, and Zerbetto}}]{ref:acocella5}
\bibinfo{author}{\bibfnamefont{A.}~\bibnamefont{Acocella}},
  \bibinfo{author}{\bibfnamefont{M.}~\bibnamefont{de~Simone}},
  \bibinfo{author}{\bibfnamefont{F.}~\bibnamefont{Evangelista}},
  \bibinfo{author}{\bibfnamefont{M.}~\bibnamefont{Coreno}},
  \bibinfo{author}{\bibfnamefont{P.}~\bibnamefont{Rudolf}}, \bibnamefont{and}
  \bibinfo{author}{\bibfnamefont{F.}~\bibnamefont{Zerbetto}},
  \bibinfo{journal}{Phys. Chem. Chem. Phys} \textbf{\bibinfo{volume}{18}},
  \bibinfo{pages}{13604} (\bibinfo{year}{2016}).

\bibitem[{\citenamefont{Gaussian09}(2009)}]{ref:gaussian09}
\bibinfo{author}{\bibfnamefont{R.~A.} \bibnamefont{Gaussian09}},
  \bibinfo{journal}{Inc., Wallingford CT}  (\bibinfo{year}{2009}).

\bibitem[{\citenamefont{Jones and Marsh}(1954)}]{ref:jones3}
\bibinfo{author}{\bibfnamefont{M.~E.} \bibnamefont{Jones}} \bibnamefont{and}
  \bibinfo{author}{\bibfnamefont{R.~E.} \bibnamefont{Marsh}},
  \bibinfo{journal}{J. Am. Chem. Soc.} \textbf{\bibinfo{volume}{76}},
  \bibinfo{pages}{1434} (\bibinfo{year}{1954}).

\bibitem[{\citenamefont{Perdew et~al.}(1996)\citenamefont{Perdew, Burke, and
  Ernzerhof}}]{ref:perdew}
\bibinfo{author}{\bibfnamefont{J.~P.} \bibnamefont{Perdew}},
  \bibinfo{author}{\bibfnamefont{K.}~\bibnamefont{Burke}}, \bibnamefont{and}
  \bibinfo{author}{\bibfnamefont{M.}~\bibnamefont{Ernzerhof}},
  \bibinfo{journal}{Phys. Rev. Lett.} \textbf{\bibinfo{volume}{77}},
  \bibinfo{pages}{3865} (\bibinfo{year}{1996}).

\bibitem[{\citenamefont{Ortenzi et~al.}(2009)\citenamefont{Ortenzi, Cappelluti,
  Benfatto, and Pietronero}}]{ortenzi_prl09}
\bibinfo{author}{\bibfnamefont{L.}~\bibnamefont{Ortenzi}},
  \bibinfo{author}{\bibfnamefont{E.}~\bibnamefont{Cappelluti}},
  \bibinfo{author}{\bibfnamefont{L.}~\bibnamefont{Benfatto}}, \bibnamefont{and}
  \bibinfo{author}{\bibfnamefont{L.}~\bibnamefont{Pietronero}},
  \bibinfo{journal}{Phys. Rev. Lett.} \textbf{\bibinfo{volume}{103}},
  \bibinfo{pages}{046404} (\bibinfo{year}{2009}).

\bibitem[{\citenamefont{Benfatto and Cappelluti}(2011)}]{benfatto_prb11}
\bibinfo{author}{\bibfnamefont{L.}~\bibnamefont{Benfatto}} \bibnamefont{and}
  \bibinfo{author}{\bibfnamefont{E.}~\bibnamefont{Cappelluti}},
  \bibinfo{journal}{Phys. Rev. B} \textbf{\bibinfo{volume}{83}},
  \bibinfo{pages}{104516} (\bibinfo{year}{2011}).

\bibitem[{\citenamefont{Marsiglio et~al.}(1988)\citenamefont{Marsiglio,
  Schossmann, and Carbotte}}]{ref:marsiglio_iterative}
\bibinfo{author}{\bibfnamefont{F.}~\bibnamefont{Marsiglio}},
  \bibinfo{author}{\bibfnamefont{M.}~\bibnamefont{Schossmann}},
  \bibnamefont{and} \bibinfo{author}{\bibfnamefont{J.~P.}
  \bibnamefont{Carbotte}}, \bibinfo{journal}{Phys. Rev. B}
  \textbf{\bibinfo{volume}{37}}, \bibinfo{pages}{4965} (\bibinfo{year}{1988}).

\bibitem[{\citenamefont{Uchiyama et~al.}(2002)\citenamefont{Uchiyama, Shen,
  Lee, Damascelli, Lu, Feng, Shen, and Tajima}}]{uchiyama2002electronic}
\bibinfo{author}{\bibfnamefont{H.}~\bibnamefont{Uchiyama}},
  \bibinfo{author}{\bibfnamefont{K.~M.} \bibnamefont{Shen}},
  \bibinfo{author}{\bibfnamefont{S.}~\bibnamefont{Lee}},
  \bibinfo{author}{\bibfnamefont{A.}~\bibnamefont{Damascelli}},
  \bibinfo{author}{\bibfnamefont{D.~H.} \bibnamefont{Lu}},
  \bibinfo{author}{\bibfnamefont{D.~L.} \bibnamefont{Feng}},
  \bibinfo{author}{\bibfnamefont{Z.-X.} \bibnamefont{Shen}}, \bibnamefont{and}
  \bibinfo{author}{\bibfnamefont{S.}~\bibnamefont{Tajima}},
  \bibinfo{journal}{Phys. Rev. Lett.} \textbf{\bibinfo{volume}{88}},
  \bibinfo{pages}{157002} (\bibinfo{year}{2002}).

\bibitem[{\citenamefont{Bohnen et~al.}(2001)\citenamefont{Bohnen, Heid, and
  Renker}}]{ref:bohnen}
\bibinfo{author}{\bibfnamefont{K.-P.} \bibnamefont{Bohnen}},
  \bibinfo{author}{\bibfnamefont{R.}~\bibnamefont{Heid}}, \bibnamefont{and}
  \bibinfo{author}{\bibfnamefont{B.}~\bibnamefont{Renker}},
  \bibinfo{journal}{Phys. Rev. Lett.} \textbf{\bibinfo{volume}{86}},
  \bibinfo{pages}{5771} (\bibinfo{year}{2001}).

\bibitem[{\citenamefont{Nicoletti et~al.}(2010)\citenamefont{Nicoletti, Limaj,
  Calvani, Rohringer, Toschi, Sangiovanni, Capone, Held, Ono, Ando
  et~al.}}]{nicoletti_prl10}
\bibinfo{author}{\bibfnamefont{D.}~\bibnamefont{Nicoletti}},
  \bibinfo{author}{\bibfnamefont{O.}~\bibnamefont{Limaj}},
  \bibinfo{author}{\bibfnamefont{P.}~\bibnamefont{Calvani}},
  \bibinfo{author}{\bibfnamefont{G.}~\bibnamefont{Rohringer}},
  \bibinfo{author}{\bibfnamefont{A.}~\bibnamefont{Toschi}},
  \bibinfo{author}{\bibfnamefont{G.}~\bibnamefont{Sangiovanni}},
  \bibinfo{author}{\bibfnamefont{M.}~\bibnamefont{Capone}},
  \bibinfo{author}{\bibfnamefont{K.}~\bibnamefont{Held}},
  \bibinfo{author}{\bibfnamefont{S.}~\bibnamefont{Ono}},
  \bibinfo{author}{\bibfnamefont{Y.}~\bibnamefont{Ando}}, \bibnamefont{et~al.},
  \bibinfo{journal}{Phys. Rev. Lett.} \textbf{\bibinfo{volume}{105}},
  \bibinfo{pages}{077002} (\bibinfo{year}{2010}).

\bibitem[{\citenamefont{Kresse and Hafner}(1993)}]{VASP1}
\bibinfo{author}{\bibfnamefont{G.}~\bibnamefont{Kresse}} \bibnamefont{and}
  \bibinfo{author}{\bibfnamefont{J.}~\bibnamefont{Hafner}},
  \bibinfo{journal}{Phys. Rev. B} \textbf{\bibinfo{volume}{48}},
  \bibinfo{pages}{13115} (\bibinfo{year}{1993}).

\bibitem[{\citenamefont{Kresse and Furthm{\"u}ller}(1996)}]{VASP2}
\bibinfo{author}{\bibfnamefont{G.}~\bibnamefont{Kresse}} \bibnamefont{and}
  \bibinfo{author}{\bibfnamefont{J.}~\bibnamefont{Furthm{\"u}ller}},
  \bibinfo{journal}{Phys. Rev. B} \textbf{\bibinfo{volume}{54}},
  \bibinfo{pages}{11169} (\bibinfo{year}{1996}).

\bibitem[{\citenamefont{Kresse and Joubert}(1999)}]{PAW}
\bibinfo{author}{\bibfnamefont{G.}~\bibnamefont{Kresse}} \bibnamefont{and}
  \bibinfo{author}{\bibfnamefont{D.}~\bibnamefont{Joubert}},
  \bibinfo{journal}{Phys. Rev. B} \textbf{\bibinfo{volume}{59}},
  \bibinfo{pages}{1758} (\bibinfo{year}{1999}).

\bibitem[{\citenamefont{Pines and Nozi{\`e}res}(1966)}]{pino66}
\bibinfo{author}{\bibfnamefont{D.}~\bibnamefont{Pines}} \bibnamefont{and}
  \bibinfo{author}{\bibfnamefont{P.}~\bibnamefont{Nozi{\`e}res}},
  \emph{\bibinfo{title}{The Theory of Quantum Liquids: Normal Fermi Liquids}},
  vol.~\bibinfo{volume}{1} (\bibinfo{publisher}{WA Benjamin},
  \bibinfo{year}{1966}).

\bibitem[{\citenamefont{Giuliani and Vignale}(2005)}]{givi08}
\bibinfo{author}{\bibfnamefont{G.}~\bibnamefont{Giuliani}} \bibnamefont{and}
  \bibinfo{author}{\bibfnamefont{G.}~\bibnamefont{Vignale}},
  \emph{\bibinfo{title}{Quantum theory of the electron liquid}}
  (\bibinfo{publisher}{Cambridge university press}, \bibinfo{year}{2005}).

\bibitem[{\citenamefont{Runge and Gross}(1984)}]{rugrprl84}
\bibinfo{author}{\bibfnamefont{E.}~\bibnamefont{Runge}} \bibnamefont{and}
  \bibinfo{author}{\bibfnamefont{E.~K.} \bibnamefont{Gross}},
  \bibinfo{journal}{Phys. Rev. Lett.} \textbf{\bibinfo{volume}{52}},
  \bibinfo{pages}{997} (\bibinfo{year}{1984}).

\bibitem[{\citenamefont{Petersilka et~al.}(1996)\citenamefont{Petersilka,
  Gossmann, and Gross}}]{pegoprl96}
\bibinfo{author}{\bibfnamefont{M.}~\bibnamefont{Petersilka}},
  \bibinfo{author}{\bibfnamefont{U.}~\bibnamefont{Gossmann}}, \bibnamefont{and}
  \bibinfo{author}{\bibfnamefont{E.}~\bibnamefont{Gross}},
  \bibinfo{journal}{Phys. Rev. Lett.} \textbf{\bibinfo{volume}{76}},
  \bibinfo{pages}{1212} (\bibinfo{year}{1996}).

\bibitem[{\citenamefont{Ceperley and Alder}(1980)}]{cealprl80}
\bibinfo{author}{\bibfnamefont{D.~M.} \bibnamefont{Ceperley}} \bibnamefont{and}
  \bibinfo{author}{\bibfnamefont{B.}~\bibnamefont{Alder}},
  \bibinfo{journal}{Phys. Rev. Lett.} \textbf{\bibinfo{volume}{45}},
  \bibinfo{pages}{566} (\bibinfo{year}{1980}).

\bibitem[{\citenamefont{Perdew and Zunger}(1981)}]{pezuprb81}
\bibinfo{author}{\bibfnamefont{J.~P.} \bibnamefont{Perdew}} \bibnamefont{and}
  \bibinfo{author}{\bibfnamefont{A.}~\bibnamefont{Zunger}},
  \bibinfo{journal}{Phys. Rev. B} \textbf{\bibinfo{volume}{23}},
  \bibinfo{pages}{5048} (\bibinfo{year}{1981}).

\bibitem[{\citenamefont{Troullier and Martins}(1991)}]{trmaprb91}
\bibinfo{author}{\bibfnamefont{N.}~\bibnamefont{Troullier}} \bibnamefont{and}
  \bibinfo{author}{\bibfnamefont{J.~L.} \bibnamefont{Martins}},
  \bibinfo{journal}{Phys. Rev. B} \textbf{\bibinfo{volume}{43}},
  \bibinfo{pages}{1993} (\bibinfo{year}{1991}).

\bibitem[{\citenamefont{Adler}(1962)}]{adpr62}
\bibinfo{author}{\bibfnamefont{S.~L.} \bibnamefont{Adler}},
  \bibinfo{journal}{Phys. Rev.} \textbf{\bibinfo{volume}{126}},
  \bibinfo{pages}{413} (\bibinfo{year}{1962}).

\bibitem[{\citenamefont{Wiser}(1963)}]{wipr63}
\bibinfo{author}{\bibfnamefont{N.}~\bibnamefont{Wiser}},
  \bibinfo{journal}{Phys. Rev.} \textbf{\bibinfo{volume}{129}},
  \bibinfo{pages}{62} (\bibinfo{year}{1963}).

\end{thebibliography}
\end{document}